\documentclass[11pt]{article}
\usepackage{soul}
\usepackage{amssymb}
\usepackage{url}
\usepackage{amsfonts}
\usepackage{amsmath}
\usepackage[top=1.2in, bottom=1.2in, left=1.2in, right=1.2in]{geometry}
\usepackage{graphicx,color,psfrag}
\usepackage{float}
\usepackage{caption}
\usepackage{subcaption}
\usepackage{hyperref}
\usepackage{multirow}
\usepackage{rotating}
\usepackage{natbib}
\usepackage{threeparttable}
\usepackage{booktabs}
\usepackage{titlesec}
\usepackage{array}
\usepackage{tikz}
\usepackage{enumitem}
\usepackage{titling}
\usepackage{bbm}
\usepackage{titling}
\usepackage{rotating}
\usepackage{authblk}
\usepackage{soul}
\usepackage{float}
\usepackage{xspace}

\usepackage{pdflscape}
\newcommand*{\figuretitle}[1]{%
	{\centering
		\footnotesize{#1}          
		\par\medskip}          
}

\hypersetup{
	pdftitle={},    
	pdfauthor={},     
	colorlinks=true,       
	linkcolor=blue,          
	citecolor=black,        
	filecolor=black,      
	urlcolor=black           
}
 
\newcommand\addtag{\refstepcounter{equation}\tag{\theequation}}
\newcolumntype{L}[1]{>{\raggedright\let\newline\\\arraybackslash\hspace{0pt}}m{#1}}
\newcolumntype{C}[1]{>{\centering\let\newline\\\arraybackslash\hspace{0pt}}m{#1}}
\newcolumntype{R}[1]{>{\raggedleft\let\newline\\\arraybackslash\hspace{0pt}}m{#1}}

\titleformat*{\section}{\centering\Large\sc}
\titleformat*{\subsection}{\large\bf}

\renewcommand{\baselinestretch}{1.3} 

\begin{document}
\title{The Influence of Neighborhood Design on the Sustainability of US Suburbs}
\author[1]{Arianna Salazar-Miranda\thanks{I am very grateful to Albert Saiz, Emily Talen, Stephen Wheeler, Eran Ben-Joseph, Adam Millard-Ball, and seminar participants at MIT's Senseable City Lab, Mansueto Institute for Urban Innovation, MIT Center for Real Estate, DUSP, and ACSP for helpful discussions and comments.}}

\affil[1]{Yale, School of the Environment, New Haven, CT, USA}

\date{October 2025}
\maketitle

\begin{abstract}
{\footnotesize
The growth of suburbs in the US has led to significant sustainability challenges; yet, it remains unclear whether these challenges stem from the remoteness of suburbs from city centers or the specific designs used to develop them. This paper examines how Garden City Design (GCD)---one of the most influential suburban design paradigms since the early 20th century---impacts the social and environmental outcomes of neighborhoods. I first introduce a composite measure of GCD, derived from street layouts and block configurations, to quantify its nationwide adoption. I use this measure combined with mobility and emissions data to estimate the impact of GCD on neighborhood outcomes using complementary identification strategies, including ordinary least squares (OLS), matching estimators, and an instrumental variables (IV) approach that exploits historical variation in GCD adoption. Results show that GCD leads to worse sustainability outcomes, including increased greenhouse gas emissions, greater social isolation, and higher sedentary behavior. The prevalence of GCD accounts for 27-38\% of the adverse effects associated with suburbanization, underscoring the crucial role that neighborhood design plays in shaping urban sustainability.

}


\end{abstract}
\thispagestyle{empty}
\setcounter{page}{0}
\clearpage
\section{Introduction}
Half of US households reside in suburban areas, making suburban development the dominant form of urbanization \citep{carroll_2002}. This widespread suburban expansion presents substantial environmental and social sustainability challenges, including increased car dependency, elevated greenhouse gas emissions, and diminished social cohesion \citep{putnam_2000, ewing_travel_2010, glaeser_2010, Wilson2010, bailey_2004, boone_2018}. 

Existing critiques of suburbanization largely emphasize macro-level factors such as distance from city centers, highlighting how sprawl exacerbates car dependency and environmental degradation \citep{newman_1989, malpezzi_1999, levinson_1997}. However, micro-level aspects of neighborhood design---such as street layouts, block configurations, and dwelling unit density---also shape residents' behaviors, influencing transportation choices, social interactions, and environmental outcomes \citep{asabere_1990, pivo_1994, cervero_1997, grannis_1998, crane_2000, krizek_2003, barrington_2017, song_2003}.

These neighborhood-level characteristics are products of planning decisions rooted in historical urban design paradigms. One particularly influential paradigm is \emph{Garden City Design} (GCD), which originated in the early 20th century in the UK and subsequently became a blueprint for suburban planning in the US \citep{jackson_crabgrass_2006, Wheeler2003, forsynth_2008, southworth_evolving_1993, wheeler_2015}. 

GCD promoted the use of winding streets, irregular block shapes, and hierarchical road systems to create aesthetically pleasing, socially cohesive, and pollution-free neighborhoods \citep{vanderryn_1986, howard_garden_1965, batty_1994}. This paradigm spread widely in the US during the twentieth century, particularly after World War II, but its prominence declined by the 1990s as critiques of its car-oriented, disconnected layouts grew, and newer paradigms, such as New Urbanism and Transit-Oriented Development, became more influential \citep{Duany2000, Talen2013}. Although less common in new developments today, GCD ideas remain embedded in many suburban neighborhoods, making it important to assess their long-term consequences.

This paper examines whether neighborhood design---particularly GCD---affects environmental and social outcomes and evaluates the magnitude of these effects relative to the influence of distance from urban centers. Understanding the relative importance of neighborhood design versus remoteness has important implications for policy. If the challenges faced by suburbs primarily stem from remoteness, policies promoting public transportation and densification around city cores should be prioritized. Conversely, if neighborhood design plays a substantial role, then strategies emphasizing design improvements in new suburban developments and targeted retrofitting of existing suburbs become a viable complement. Recent initiatives, such as the Infrastructure Investment Act, support the removal of highways, reconnecting neighborhoods, and investing in pedestrian and cycling infrastructure, signaling a renewed policy interest in suburban retrofitting.

This study makes two primary contributions. First, it introduces a composite measure of GCD's influence, derived from street network and block structure attributes, which captures the defining features of this planning paradigm. Prior research predominantly examined isolated attributes such as street connectivity or density \citep[see, for example,][]{moudon_built_1989, march_geometry_1971, hillier_natural_1993, Wheeler2003, hillier_social_2005, Barthelemy2011, strano_2012, marshall_streets_2005, barthelemy2014a}. Instead, this composite measure explicitly recognizes that these attributes originate in the GCD paradigm and, therefore, did not evolve independently or at random. The composite measure facilitates the systematic tracing of GCD adoption and directly links contemporary neighborhood designs---and their impacts---to their historical planning origins. 

Additionally, this composite measure enables an analysis of GCD's influence at a national scale. This represents a significant shift from previous approaches that focused on smaller geographic areas or manual map inspections to identify the influence of different design paradigms \citep[as in work by][]{conzen_thinking_2004, strano_2012, Wheeler2003}. 

Using this measure, the study documents three key facts: 1) GCD adoption began in the early 1900s, increased significantly after World War II, peaked around 1990, and has declined since then; 2) the prominence of GCD is not merely a byproduct of suburban sprawl; its adoption waxed and waned in time, both near and far from city centers; and 3) GCD adoption is better understood as a nationwide phenomenon, largely independent of local geographic or topographic constraints. 

The second contribution of this study is to evaluate the social and environmental consequences of GCD adoption. Prior research has explored correlations between specific design elements and behavioral or transportation outcomes \citep[see][]{cervero_1997, grannis_1998, cao_2010, hajrasouliha_2015, barrington_2017, barrington2019global, boeing_2020}. This study offers a new perspective by assessing the GCD paradigm as a whole and its effects on neighborhood sustainability across the US. It examines three key outcomes---social isolation, sedentarism, and greenhouse gas emissions---capturing how design influences both social behavior and environmental impact. Social outcomes are derived from GPS-based mobility data, and environmental outcomes are obtained from the Environmental Protection Agency (EPA).

To estimate the impact of GCD, the study uses three complementary identification strategies. The first approach relies on OLS estimates, exploiting variation in GCD adoption across neighborhoods while controlling for a range of characteristics, including geography and location. The second strategy matches neighborhoods that adopted GCD with observationally similar neighborhoods that did not, ensuring that differences in outcomes can be attributed specifically to GCD rather than underlying geographic differences. The third strategy employs an Instrumental Variable (IV) approach that leverages variations in neighborhood design resulting from historical design waves during which GCD gained popularity. 


\section{Results}\label{results}

\subsection{Measuring the influence of GCD}\label{results}

To quantify the adoption of Garden City Design (GCD) across the United States, I construct a composite GCD index for 60,421 \textit{urban} neighborhoods. These are defined as Census Block Groups with more than half their area within urbanized boundaries (see Methods Section \ref{section:neigh definition} for sample details). 

The index combines four design features that planning historians identify as central to the Garden City model: (1) organic block layouts with irregular shapes, (2) hierarchical street networks separating major and minor roads, (3) enclosed streets such as cul-de-sacs, and (4) curvilinear street patterns emphasizing winding routes. These features represent the physical form through which GCD sought to achieve its aesthetic, social, and environmental ideals \citep[see][]{cleveland_1871, creese_1966, stern_paradise_2013, robinson_1926}. 


The four components are measured using data from OpenStreetMap \citep{openstreetmap_2021}, which describes the street layouts and block shapes of all US neighborhoods. Each component is calculated separately (as described in Methods Section \ref{section:GCD construction}), standardized, and averaged into a single GCD composite index, which is then re-centered to range from 0 to 1. Higher index values indicate stronger adherence to Garden City principles. The four components are given equal weights to capture the \textit{bundle} of design traits emphasized by the paradigm. The results are robust to alternative weighting schemes such as principal component analysis. 

To validate the GCD index, I compare it against three independent historical classifications of neighborhoods known to exemplify Garden City principles, from previous work by \citet{wheeler_evolution_2008}, \citet{ SalazarMiranda2021} and \citet{talen_2022} for a subset of US locations. The details of this validation exercise are provided in Methods Section \ref{section:GCD construction}. All neighborhoods classified as exemplars of Garden City Design fall in the top quintile of the GCD index distribution. This shows that the composite measure successfully identifies canonical garden city developments.

\subsubsection{The Adoption of GCD across time and space}

\begin{figure}[!ht]
	\centering
	\includegraphics[width=\linewidth]{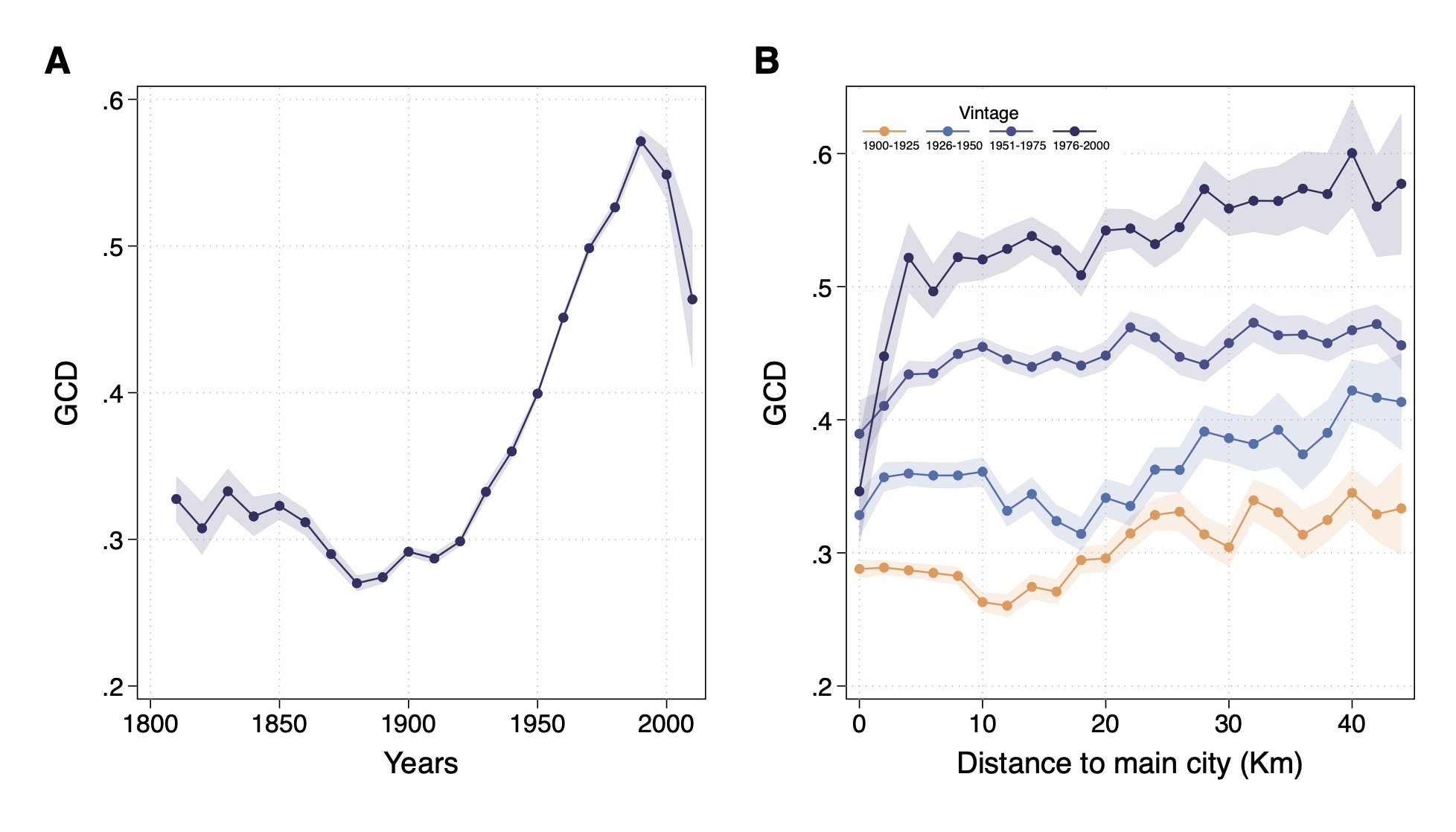}
	
\caption{\small \textsc{Temporal and Spatial Dynamics of GCD.} Panel A illustrates GCD trends in 5-year intervals from 1800 to 2015. Panel B plots the relationship between GCD and urban sprawl, measured as distance to the primary city within each MSA, for neighborhoods established during four distinct periods: 1900–1925, 1926–1950, 1951–1975, and 1976–2000. For visual clarity, the graph focuses on neighborhoods within 45 km of city centers, capturing 90\% of the sample. The error bands represent 95\% confidence intervals.}
	\label{fig: gdi time}
\end{figure}

I then use the composite index to examine how GCD spread across time and regions. Figure \ref{fig: gdi time} Panel A, plots the average GCD index by neighborhood development year. The development year is defined as the modal year of construction within each neighborhood, using data from \citet{leyk_2018} on the earliest recorded building date at a 250-meter resolution. For example, if the majority of pixels in a neighborhood were first developed in 1950, that neighborhood is classified as developed in 1950.

Neighborhoods developed during the nineteenth century were predominantly grid- and streetcar-based, showing little influence of Garden City principles. GCD adoption rose sharply in the early twentieth century following its introduction from the UK, peaked around 1990, and declined thereafter. Its rise coincided with the postwar suburban expansion, when large-scale private development---shaped by local political, social, and economic conditions---facilitated the widespread diffusion of GCD layouts \citep{hayden_2003}. After 1990, the paradigm lost prominence as interest shifted back toward grid-based and compact development patterns \citep{HAMIDI201472, Barrington-Leigh2015, boeing_2020}.

Figure \ref{fig: gdi time} Panel B examines how GCD adoption varies with urban sprawl across four periods: 1900-1925, 1926-1950, 1951-1975, and 1976-2000. Sprawl is measured as the distance from each neighborhood to the most populous city center in its MSA.  Across all four periods, neighborhoods farther from city centers consistently show higher GCD values. For example, among those established between 1951 and 1975, neighborhoods close to city centers have an average GCD index of 0.40, compared with 0.46 for those located 40 km away. 

Importantly, the findings in Panel B show that the increase in GCD over time is not simply a byproduct of sprawl. Adoption rose at all distances from city centers over time. At 5 km, average GCD indices increased from 0.27 in early cohorts to 0.50 by the 1990s; at 45 km, they rose from 0.34 to 0.57. While GCD was more popular in suburban locations, its expansion across all distances reflects a national design trend that reshaped both urban and suburban neighborhoods. This fact will later support the identification strategy, since it shows that variation in neighborhood design arose not only from outward expansion but also from nationwide design trends.

\begin{figure}[!ht]
	\centering
	\includegraphics[width=\linewidth]{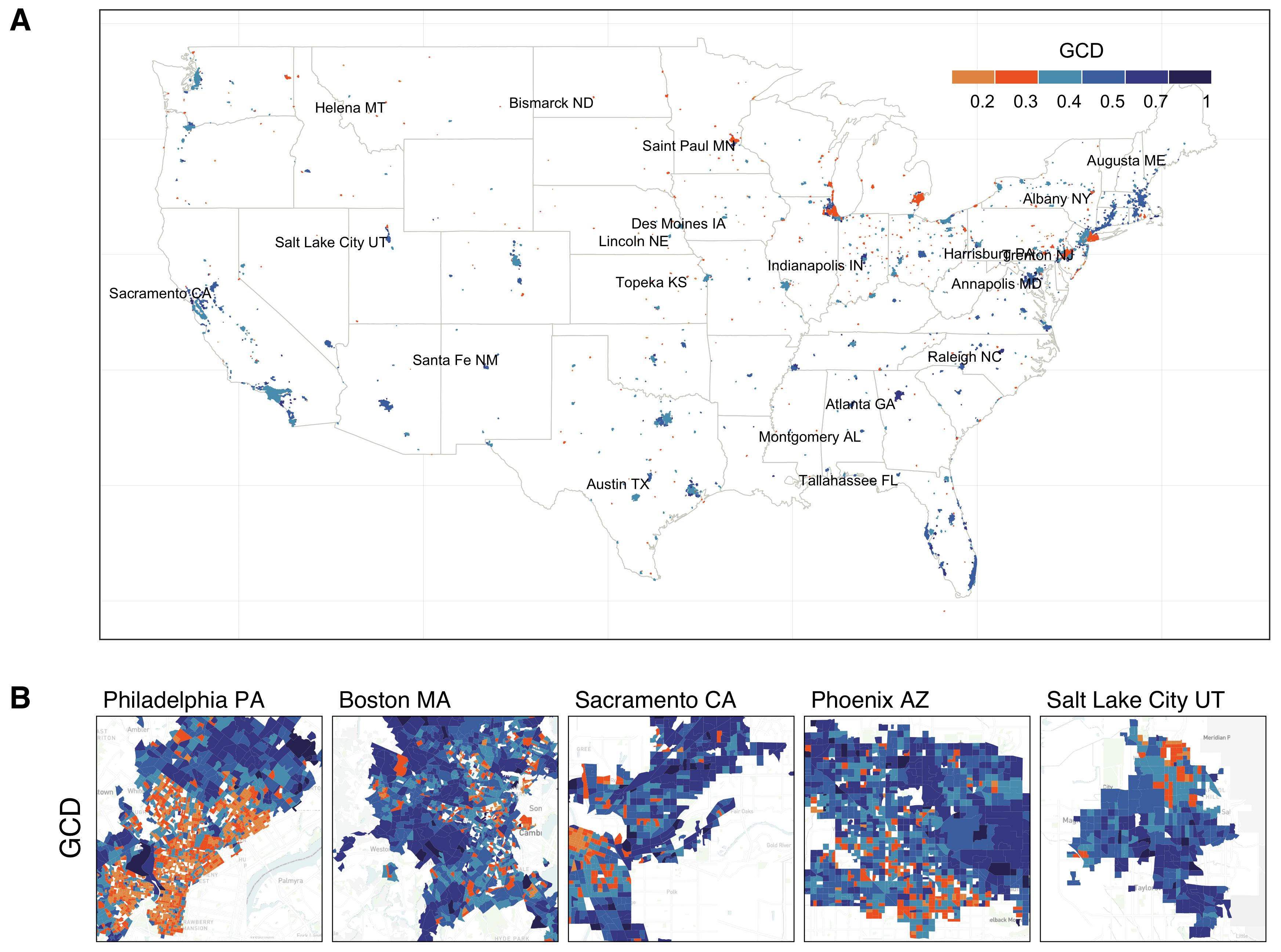}
	
\caption{\small \textsc{Variation of GCD in Urban Areas.} Panel A illustrates the average GCD index across urban areas (N=22,494). Panel B shows the GCD index for neighborhoods in five selected urban areas: Philadelphia, Boston, Sacramento, Phoenix, and Salt Lake City.}
	\label{fig: gdi maps}
\end{figure}

Figure \ref{fig: gdi maps} shows the spatial distribution of GCD across US urban areas. Neighborhoods with mid-to-high GCD indices appear in nearly every urban area nationwide. Only 13.6\% of the total variation in GCD is explained by differences \textit{across} urban areas, while 86.4\% reflects variation \textit{within} them. Panel B highlights this within-area heterogeneity in five urban areas---Philadelphia, Boston, Sacramento, Phoenix, and Salt Lake City. Within the same urban area, neighborhoods often diverge sharply in their design, with some exhibiting high GCD features and others more regular grid-based layouts.

\subsection{Social and Environmental Impacts of GCD:}\label{descriptive results}

This section examines how GCD influences three neighborhood-level outcomes central to sustainability: social isolation, sedentarism, and greenhouse gas (GHG) emissions associated with car travel. Below, I briefly describe the data sources and construction of these outcomes. Further methodological details are provided in Methods Section \ref{section:outcome construction}. 

Social isolation is measured using SafeGraph data (2019), which track monthly trips from nearly 40 million smartphone users to points of interest (POIs), such as parks, schools, and commercial establishments. For each neighborhood, I identify all POIs within its boundaries and calculate two types of visits: those made by residents (local visits) and those made by non-residents (external visits). I then compute a standard exposure measure of residents to non-residents, defined as the product of local visits per resident (the frequency of visits) and the share of non-residents among visitors (the probability of exposure to non-residents during these visits). This measure therefore approximates how often residents are exposed to people from outside their neighborhood when visiting nearby amenities. Prior research shows that such spatial proximity strongly predicts social tie formation \citep{Festinger1950, SmallAdler2019}. To express this as a measure of isolation, I take the negative logarithm of the exposure measure, so that proportional differences are comparable across neighborhoods and higher values indicate greater social isolation (fewer interaction opportunities).

Sedentarism is measured as the average daily minutes residents spend at their home location during waking hours. This measure also uses SafeGraph data (2019). It identifies each smartphone user's primary residence based on their location history and calculates the total daily minutes spent at this location during waking hours. This measure proxies for sedentarism, as greater wake time spent at home typically corresponds to fewer daily activities and reduced physical movement \cite{mccarthy_2021, DeOliveiradaSilvaScaranni2023}.

GHG emissions are measured using data from the Environmental Protection Agency (EPA, 2017), which reports data on emissions and vehicle miles traveled (VMT) by residents of each US neighborhood. The EPA estimates daily VMTs per person for both commute and non-commute trips using data from the National Household Travel Survey (NHTS). I annualize these two measures (assuming 260 commute days and 365 days for regular travel), add them, and then estimate emissions by applying a standard EPA conversion factor of 0.0004 metric tons of CO$_2$ per mile traveled. This conversion factor represents the average emissions intensity of the US vehicle fleet, taking into account average fuel types and vehicle models.  

To examine how these outcomes vary with neighborhood design, I compare neighborhoods classified into two groups based on their GCD index:  neighborhoods in the top 20\% of the distribution (``high GCD'') and those in the bottom 80\% (``low GCD''). This threshold choice is supported by the fact that historically recognized Garden City developments consistently fall within or near the top quintile of the GCD distribution (as discussed in the validation and shown in Figure \ref{fig: gdi validation}). Results are similar when using alternative definitions, such as the top tertile of the GCD, the continuous GCD index, or the first principal component of the four Garden City Design features (see Appendix Tables \ref{table: all continuous}–\ref{table: tertile robust}). 

Table \ref{table: summary statistics} summarizes the outcomes for the full sample and separately for high-and low-GCD neighborhoods. Panel I shows that high GCD neighborhoods have higher social isolation (logged value of 3.62) than those with low GCD (logged value of 3.43). This 19-log-point difference corresponds to a 20\% increase in social isolation ($\exp(0.19)$). Residents of high-GCD neighborhoods also spend more time at home---661 minutes per day---compared to 640 in low-GCD neighborhoods. Annual per capita GHG emissions from car trips are also higher in high-GCD neighborhoods (2.5 metric tons per person) compared to low-GCD ones (2.15 metric tons per person). 

\begin{table}[!ht]
	\centering
	\caption{\sc{Summary Statistics}}
	\label{table: summary statistics}
	\resizebox{\textwidth}{!}{\begin{tabular}{L{8cm}C{4cm}C{4cm}C{4cm}}\toprule\toprule
			&\multicolumn{3}{c}{\sc{Samples}}\\\cmidrule(r){2-4}
			\vspace{0.2cm}
			&\multicolumn{1}{c}{All Sample} & 	\multicolumn{1}{c}{High-GCD} & 	\multicolumn{1}{c}{Low-GCD}  \\
			&\multicolumn{1}{c}{(N = 60,421)} & 	\multicolumn{1}{c}{(N = 12,084)} & 	\multicolumn{1}{c}{(N = 48,337)}  \\			
			\\
			&\multicolumn{3}{c}{\sc{Panel I. Outcomes}}\\\cmidrule(r){2-4}
			\multirow{2}{7cm}{Log social isolation\dotfill}&    3.463 &     3.618 &     3.426 \\   
			& (    1.199) &  (    1.337) &  (    1.160) \\ 
			\multirow{2}{7cm}{Daily at-home time\dotfill}&  644.449 &   661.319 &   640.231 \\   
			& (   85.679) &  (   87.110) &  (   84.796) \\ 
			\multirow{2}{7cm}{Annual ghg per person\dotfill}&    2.220 &     2.489 &     2.152 \\   
			& (    0.765) &  (    0.726) &  (    0.760) \\ 
			\\
			&\multicolumn{3}{c}{\sc{Panel II. Controls}}\\\cmidrule(r){2-4}
            \multirow{2}{7cm}{Distance to City Center (km)\dotfill}&   21.822 &    26.683 &    20.607 \\   
			& (   21.539) &  (   21.031) &  (   21.493) \\ 
			\multirow{2}{7cm}{Elevation\dotfill}&  199.670 &   216.548 &   195.451 \\   
			& (  329.065) &  (  367.456) &  (  318.610) \\ 
			\multirow{2}{7cm}{Slope\dotfill}&    0.508 &     0.602 &     0.484 \\   
			& (    0.582) &  (    0.691) &  (    0.549) \\ 
			\multirow{2}{7cm}{Latitude\dotfill}&   37.922 &    36.854 &    38.189 \\   
			& (    4.889) &  (    5.116) &  (    4.794) \\ 
			\multirow{2}{7cm}{Longitude\dotfill}&  -93.883 &   -95.764 &   -93.413 \\   
			& (   17.946) &  (   18.585) &  (   17.752) \\ 
            \multirow{2}{7cm}{Share in Midwest\dotfill}&    0.208 &     0.119 &     0.230 \\   
			& (    0.406) &  (    0.324) &  (    0.421) \\ 
			\multirow{2}{7cm}{Share in Northeast\dotfill}&    0.237 &     0.189 &     0.249 \\   
			& (    0.425) &  (    0.391) &  (    0.433) \\ 
			\multirow{2}{7cm}{Share in South\dotfill}&    0.230 &     0.302 &     0.212 \\   
			& (    0.421) &  (    0.459) &  (    0.408) \\ 			
			\multirow{2}{7cm}{Share in West\dotfill}&    0.325 &     0.390 &     0.309 \\   
			& (    0.468) &  (    0.488) &  (    0.462) \\ 
			\\\bottomrule
	\end{tabular}}
	\begin{minipage}{1\linewidth}											
		\scriptsize \textsl{Note.—Columns 1–3 report the mean and standard deviation (in square brackets) for each variable. Column 1 reports these for the sample of urban neighborhoods (N=60,421). Column 2 is for neighborhoods classified as high GCD (top 20\% in the GCD distribution, N = 12,084), and column 3 is for low-GCD neighborhoods (remaining 80\%, N = 48,337). Panel I reports the primary outcomes analyzed in the study: social isolation; daily at home time during waking hours; and annual GHG emissions from residents' travel behavior. Panel II provides descriptive statistics for covariates used in the analysis.}	 
	\end{minipage}	
\end{table}	

The main challenge in interpreting these differences as evidence for a causal effect of design is that high-GCD neighborhoods differ along other dimensions. For example, Panel II of Table~\ref{table: summary statistics} shows that high-GCD neighborhoods are typically farther away from city centers, occupy more rugged terrains with steeper slopes, and are more likely to be located in certain US regions. These differences underscore the need for credible identification strategies when investigating the impact of neighborhood design. I rely on three complementary identification strategies: OLS controlling for geographic differences, propensity score matching, and instrumental variables (IV).

\subsubsection*{OLS with covariates:}\label{ols}

The first strategy estimates the relationship between GCD and neighborhood outcomes via OLS, using the following specification:
\begin{align*}
\textrm{Outcome}_{ims} = &
\beta\;\textrm{GCD}_{ims} + F(d(i)) + \alpha_{m} + \gamma_{s} + \theta\; X_{ims} + \epsilon_{ims}.
\addtag\label{eq:ols}
\end{align*}
The unit of analysis is a neighborhood $i$ in metropolitan area $m$ and state $s$. Here, $\textrm{GCD}_{ims}$ is a dummy indicating whether the neighborhood is in the high-GCD category. $\beta$ captures the association between GCD and each outcome. The model includes controls for distance to the main city center in the MSA (denoted by $d(i)$), using dummies for 5-km distance bins. It also includes metropolitan area and state fixed effects ($\alpha_{m}$ and $\gamma_{s}$) and a set of geographic covariates ($X_{i}$), such as elevation, slope, ecological region dummies, latitude, and longitude.  $\epsilon_{ims}$ is the error term, assumed orthogonal to $\textrm{GCD}_{ims}$.

The full set of controls ensures that the coefficient of interest, $\beta$, is identified from differences in GCD design among neighborhoods within the same MSA and state that lie at similar distances from the city center and share comparable geographic characteristics. A causal interpretation of $\beta$ assumes that, conditional on these geographic and locational controls, the remaining variation in GCD adoption reflects exogenous historical factors---such as the preferences of individual developers---rather than unobserved neighborhood attributes that independently affect the outcomes of interest and are captures by the error term $\epsilon_{ims}$. This assumption is plausible because the spread of GCD in the US was largely driven by design fashions popular among developers at the time, rather than by systematic local conditions \citep[see, for example,][for some historical support]{hayden_2003}.

Table \ref{table: all discrete} presents OLS estimates, with standard errors clustered at the county level to account for spatial correlation across neighborhoods. Panel I reports estimates for social isolation as the dependent variable, Panel II for daily wake-time at home, and Panel III for annual GHG emissions. Column 1 controls for location using 5-km distance bins to the main city center. Column 2 includes all covariates listed in equation \eqref{eq:ols}. 

The main OLS estimates in column 2 imply that moving from a low- to high-GCD neighborhood increases social isolation by 19 log points (a 20\% reduction in the probability of encountering non-residents), raises daily wake-time at home by 15 minutes, and increases annual per-capita GHG emissions from car use by 0.185 metric tons.

\begin{table}[!ht]
	\centering
	\caption{\sc{Estimates of Garden Design on Social and Environmental Outcomes}}
    \label{table: all discrete}
	\resizebox{1\textwidth}{!}{\begin{tabular}{L{6.5cm}C{2cm}C{2cm}C{2cm}C{2cm}C{2cm}}\toprule\toprule
			\vspace{0.5cm}
			&(I) &(II)&(III)&(IV)&(V)\\
			\cmidrule(r){2-3} \cmidrule(r){4-4}  \cmidrule(r){5-6} \\
			&\multicolumn{2}{c}{OLS Estimates} & 	\multicolumn{1}{c}{Propensity Score} & 	\multicolumn{2}{c}{IV Estimates}   \\
			\vspace{0.5cm}
			&\multicolumn{5}{c}{\sc{Panel I. Dependent Variable:  Log Social isolation}}\\\cmidrule(r){2-6}	
			        &                    &                    &                    &                    &                    \\
High-GCD neighborhoods (Top 20\%)&       0.207$^{***}$&       0.192$^{***}$&       0.166$^{***}$&       0.375$^{***}$&       0.403$^{***}$\\
            &     (0.026)        &     (0.026)        &     (0.026)        &     (0.072)        &     (0.066)        \\
Observations&       45226        &       45226        &       45226        &       45226        &       45226        \\
R-squared   &        0.01        &        0.05        &                    &        0.01        &        0.02        \\
F-Stat      &                    &                    &                    &      551.73        &     1242.11        \\

			\vspace{0.2cm}
			&\multicolumn{5}{c}{\sc{Panel II. Dependent Variable:  Daily Time at Home (minutes)}}\\\cmidrule(r){2-6}	
			       &                    &                    &                    &                    &                    \\
High-GCD neighborhoods (Top 20\%)&      12.920$^{***}$&      15.052$^{***}$&      14.755$^{***}$&      58.133$^{***}$&      64.367$^{***}$\\
            &     (3.086)        &     (1.671)        &     (1.865)        &    (10.504)        &     (5.553)        \\
Observations&       60348        &       60348        &       60348        &       60348        &       60348        \\
R-squared   &        0.07        &        0.18        &                    &        0.01        &        0.01        \\
F-Stat      &                    &                    &                    &      671.96        &     1697.81        \\

			\vspace{0.2cm}
			&\multicolumn{5}{c}{\sc{Panel III. Dependent Variable:  Annual GHG (metric tons)}}\\\cmidrule(r){2-6}
			        &                    &                    &                    &                    &                    \\
High-GCD neighborhoods (Top 20\%)&       0.344$^{***}$&       0.185$^{***}$&       0.238$^{***}$&       1.279$^{***}$&       0.519$^{***}$\\
            &     (0.054)        &     (0.033)        &     (0.028)        &     (0.178)        &     (0.108)        \\
Observations&       60353        &       60353        &       60353        &       60353        &       60353        \\
R-squared   &        0.07        &        0.50        &                    &       -0.19        &        0.14        \\
F-Stat      &                    &                    &                    &      671.98        &     1697.57        \\
	
			\\
			\textsl{Controls:}\\
			Distance &  \checkmark & \checkmark & \checkmark & \checkmark & \checkmark\\
			Geography &  & \checkmark & \checkmark & & \checkmark \\
			State \& Metro Fixed Effects &  & \checkmark  & \checkmark & & \checkmark \\
			\\\bottomrule
	\end{tabular}}
	\begin{minipage}{1\linewidth}												
		\scriptsize \textsl{Note.--- Columns 1-2 present OLS estimates of the relationship between GCD (using the discrete measure) and neighborhood outcomes. Column 1 controls for distance to city center, using 5-km dummies. Column 2 adds geographic controls, including elevation, slope, soil type (ecological regions), latitude and longitude, and metro area and state fixed effects. Column 3 reports estimates from propensity score matching, using the same set of covariates as Column 2 to estimate the propensity score. Columns 4–5 provide IV estimates of the specifications from Columns 1–2, using historical cohorts as instruments for GCD adoption. The coefficients with *** are significant at the 1\% confidence level; with ** are significant at the 5\% confidence level; and with * are significant at the 10\% confidence level. Robust standard errors, adjusted for clustering by county, are in parentheses. OLS and IV estimates are weighted by neighborhood population.}			
	\end{minipage}	
\end{table}			

Although the estimates in Table \ref{table: all discrete} control for a comprehensive set of covariates, one might still worry that the estimated relationship between GCD and outcomes may be confounded by demographic or historical differences across neighborhoods. The Appendix addresses these concerns through additional robustness checks. The first test examines whether historically low-density neighborhoods were more likely to adopt GCD, in which case the OLS estimates could reflect the lingering effects of earlier development patterns. Table \ref{table: ols discrete robustness} adds controls for building density in 1900---before GCD became common---and shows that the estimates remain qualitatively similar after accounting for these historical factors.  

A second set of tests examines whether the results are driven by contemporary demographic differences. For example, high-GCD neighborhoods could be more prevalent in affluent or predominantly white areas. Table \ref{table: ols discrete robustness} addresses this by including controls for income, education, racial composition, age, and housing structure using 2000 Census data. The results remain similar after including these covariates. The  last column of Table \ref{table: ols discrete robustness} also examines whether differences in sedentarism might partly reflect remote work rather than physical inactivity. Controlling for the share of county-level workers who reported working from home (ACS 2015–2019) produces comparable estimates.

The Appendix also provides a complementary decomposition of GCD’s effect on vehicle emissions, distinguishing between commute and non-commute trips. Table \ref{table: complementary outcomes} shows that three-quarters of the increase in emissions is attributable to commute trips, and one-quarter to non-commute trips. Commute-related emissions may increase because GCD neighborhoods have limited local jobs or street layouts that make them difficult to enter and leave without a car. Non-commute emissions, in contrast, may increase because high-GCD neighborhoods tend to have fewer points of interest and a less walkable environment. Consistent with this interpretation, the same table shows that high-GCD neighborhoods have higher shares of car ownership, fewer POIs, and lower walkability scores. 

\subsubsection*{Propensity Score Matching:}\label{matching}

OLS estimates control for location and geographic differences between neighborhoods using the linear model in Equation \eqref{eq:ols}. The second identification strategy controls for these differences using propensity score matching. 

The identification assumption is similar to that of OLS: conditional on the covariates described above, whether a neighborhood was developed with GCD is independent of its potential outcomes. The difference is that propensity score matching estimates the average treatment effect of GCD by comparing neighborhoods with similar propensity scores (probabilities) of adopting GCD, based on their geographic and locational attributes. Intuitively, when two neighborhoods have identical propensity scores, whether they were built with GCD or not is as good as random. Any differences in outcomes between them can therefore be attributed to design rather than underlying geography or demographics. 

I estimate the average treatment effect using inverse propensity score reweighting, following \citet{Hirano2003}. Propensity scores are estimated via a Probit model using the covariates from Column 2 of Table~\ref{table: all discrete}. Figure \ref{fig: propensity score} in the Methods Section shows substantial overlap in estimated propensity scores between treated and control neighborhoods, satisfying the overlap condition required for this approach. The average treatment effect is computed by comparing outcomes between treated and control neighborhoods after reweighing the sample. This reweighting assigns more weight to neighborhoods where GCD adoption deviates most from what would be predicted based on their observed characteristics---that is, where adoption appears more random. Appendix Table \ref{table: covariate balance} shows that this procedure successfully balances the covariates, with no statistically significant differences in geography or location between treated and control neighborhoods. 

The results are shown in Column 3 of Table~\ref{table: all discrete}. Residents of high-GCD neighborhoods are 16.6 log points more socially isolated (18\% less likely to encounter non-residents), spend 15 additional minutes per day at home, and emit 0.24 more annual metric tons of GHG emissions per capita than residents of low-GCD neighborhoods. These estimates are similar to the OLS counterparts in Column 2, indicating that the findings are robust to controlling for key covariates either linearly (as in OLS) or through propensity score reweighing.

\subsubsection*{Instrumental Variables:}\label{iv}

The validity of the OLS and propensity score estimates depends on the assumption that, conditional on the rich set of covariates, the adoption of GCD is driven by exogenous factors. The third strategy leverages national design waves---shown in Panel A of Figure \ref{fig: gdi time}---as a plausible source of exogenous variation in GCD adoption. These waves reflect the nationwide rise in the popularity of GCD between 1940 and 1990, followed by its decline thereafter. 

Formally, I estimate Equation \ref{eq:ols}, instrumenting $\textrm{GCD}_{ims}$ with $\overline{\textrm{GCD}}(t_i)$, the national average level of GCD among neighborhoods developed in the same five-year period $t_i$ as neighborhood $i$. The instrument captures how prevalent GCD was nationally when each neighborhood was developed. The underlying intuition is that neighborhoods developed during periods when GCD was especially popular were more likely to adopt its defining design elements---not because of local geography or demographics, but because those were the prevailing planning fashions shaping developers' decisions at the time. These national waves provide exogenous variation in neighborhood design, allowing me to identify the causal effects of GCD.

The IV estimates are reported in Columns 4-5 of Table~\ref{table: all discrete} for all three main outcomes. Column 4 includes controls for distance to the city center, while Column 5 adds geographic covariates and MSA and state fixed effects. The estimates in Column 5 imply that moving a neighborhood from low to high GCD increases social isolation by 40 log points (a 50\% decrease in the probability of encountering non-residents), raises daily wake-home time at home by 64 minutes, and increases annual per capita GHG emissions by 0.52 metric tons. 

The IV estimates are roughly two to four times larger than their OLS counterparts. Three factors may account for this difference. First, measurement error in the GCD index attenuates the OLS estimates. Second, the IV captures the full bundle of design features introduced during each GCD wave, including aspects that the composite index may only partially reflect. Third, IV estimates identify a local average treatment effect (LATE), which can differ from OLS estimates if design has a stronger impact in \emph{complier} neighborhoods (i.e., those that adopted GCD because it was fashionable at the time but would not have done so otherwise).

For the IV estimates to be valid, two conditions must hold: \textit{relevance}, meaning the instrument predicts GCD adoption, and \textit{exclusion}, meaning that national design waves affect outcomes only through GCD and not other channels. Specifically, the exclusion restriction requires that national design waves are orthogonal to the residual term $\varepsilon_{ims}$ in Equation \eqref{eq:ols}.

The relevance condition is well supported by the data. Panel A of Figure Figure \ref{fig: gdi time} shows that  national waves predict adoption of GCD. This graphical evidence is confirmed by the first-stage regression results. Across all specifications in Table \ref{table: all discrete}, the first-stage $F-$ statistic exceeds 500, indicating the high explanatory power of the instrument and a strong link between national waves and local design adoption. Table \ref{table: variance decomposition} in the Appendix provides further support, showing that design cohorts explain 20\%--29\% of the variation in GCD adoption, compared to only 5--12\% 
explained by geography and other local attributes. 

While the exclusion restriction cannot be tested directly, several threats can be ruled out. One possible concern is that national GCD waves may coincide with other long-term trends in planning or economic development that independently affect neighborhood outcomes. For example, the rise of  GCD in the 1940s could overlap with the expansion of residential zoning or the spread of car-oriented suburban growth. Historical evidence, however, suggests that these overlaps are limited and do not affect the IV estimates. Appendix Figure~\ref{fig: zoning gdi time} shows that residential zoning followed a distinct historical trajectory, peaking well before GCD's widespread adoption. This temporal mismatch reduces concerns that zoning policies drive the observed relationship between GCD and neighborhood outcomes. In addition, Appendix Table \ref{table: zoning} reports similar IV estimates after controlling for the share of residentially zoned parcels in each neighborhood (see Methods Section \ref{section:data sources} for details on the zoning data).

To further account for slow-moving changes in urban planning practices, such as the rise of automobile dependence, I control for broad development periods. Appendix Table \ref{table: iv robustness} adds dummies for three eras: 1810–1875 (grid-based, pre-GCD), 1875–1950 (early GCD and industrial expansion), and 1950–2020 (postwar suburbanization and automobile dependence). These controls ensure that identification relies on variation within periods rather than across them. The IV estimates remain stable and precisely estimated after including these controls. 

A second potential is that neighborhoods developed during GCD's peak popularity (1950-1990) are younger and have had less time to accumulate social or infrastructural capital. These \emph{age effects} could, in principle, invalidate the exclusion restriction if newer neighborhoods systematically differ from older ones in ways unrelated to design. To evaluate this possibility, I first estimate the relationship between neighborhood age and outcomes using the sample of neighborhoods developed before 1875—a period with minimal design variation \citep[see][and Figure \ref{fig: gdi time}]{wheeler_evolution_2008}. I then re-estimate the IV specification for all neighborhoods after adjusting outcomes for these estimated age effects. Appendix Table \ref{table: age effects} shows that the adjusted estimates remain positive and precisely estimated, indicating that the main findings are not driven by differences in accumulated social or infrastructural capital. Details of this procedure are provided in Methods Section \ref{section: estimation}.

A final concern is that state-level factors---such as infrastructure investments or developer incentives---may have jointly influenced both the adoption of GCD and neighborhood outcomes. If so, the national GCD prevalence used as an instrument could partly reflect state-specific policies rather than independent national design waves. To test this, Appendix Table \ref{table: iv robustness} presents results using a  \textit{leave-one-out} version of the instrument, constructed as the average prevalence of GCD in all other states during the same development period. This approach captures nationwide trends while excluding state-level influences. The resulting estimates are very similar to the baseline IV results, suggesting that findings are not driven by state-specific policies or local planning environments. 

\subsubsection*{Behavioral responses or sorting?}\label{matching}

The results support the hypothesis that GCD negatively affects social and environmental outcomes. One interpretation is that the physical design elements of GCD directly influence resident behavior. For instance, irregular blocks restrict parcel subdivision and limit opportunities for commercial activity \citep[see][]{conzen_alnwick_1960, siksna_city_1998, ogrady_spatial_2014}. Similarly, curvilinear street layouts lengthen pedestrian routes and reduce convenient access to amenities, thereby lowering walkability and increasing car dependency \citep[see][]{cervero_1997, barrington_2017, hajrasouliha_2015, SalazarMiranda2021}.

An alternative explanation is residential sorting---high-GCD neighborhoods may attract residents who already behave differently, rather than design itself shaping behavior. Distinguishing between these interpretations is important: if behavior is induced by design, physical retrofits can improve sustainability; if outcomes reflect sorting, reforms would simply redistribute existing behaviors spatially, without reducing emissions or social interaction overall.

To test this, I re-estimate the main specifications using only neighborhoods located in counties with low migration rates---those in the bottom half of the national distribution---based on IRS migration data. If sorting were the primary driver, the estimated effects of GCD should weaken in this low-mobility sample. Appendix Table~\ref{table: low migration} shows that the point estimates remain similar in magnitude to those in Table~\ref{table: all discrete}, suggesting that GCD primarily shapes outcomes by influencing  behavior rather than sorting.

\subsection{Quantifying the Costs of GCD}

The previous section showed that GCD is associated with poorer neighborhood outcomes. This section quantifies the magnitude of those costs and compares them to the broader effects of suburban sprawl.

To do so, I re-estimate the effects of GCD separately for neighborhoods grouped into 10-km rolling windows by distance from the city center, using the IV specification reported in Column 5 of Table \ref{table: all discrete}. Estimating the model within distance bands allows the influence of GCD to vary with remoteness. This captures the possibility that neighborhood design plays a smaller role in outlying suburbs, where distance itself is already the dominant constraint.

\begin{figure}[!ht]
	\centering
	\figuretitle{}
	\includegraphics[width=1\linewidth]{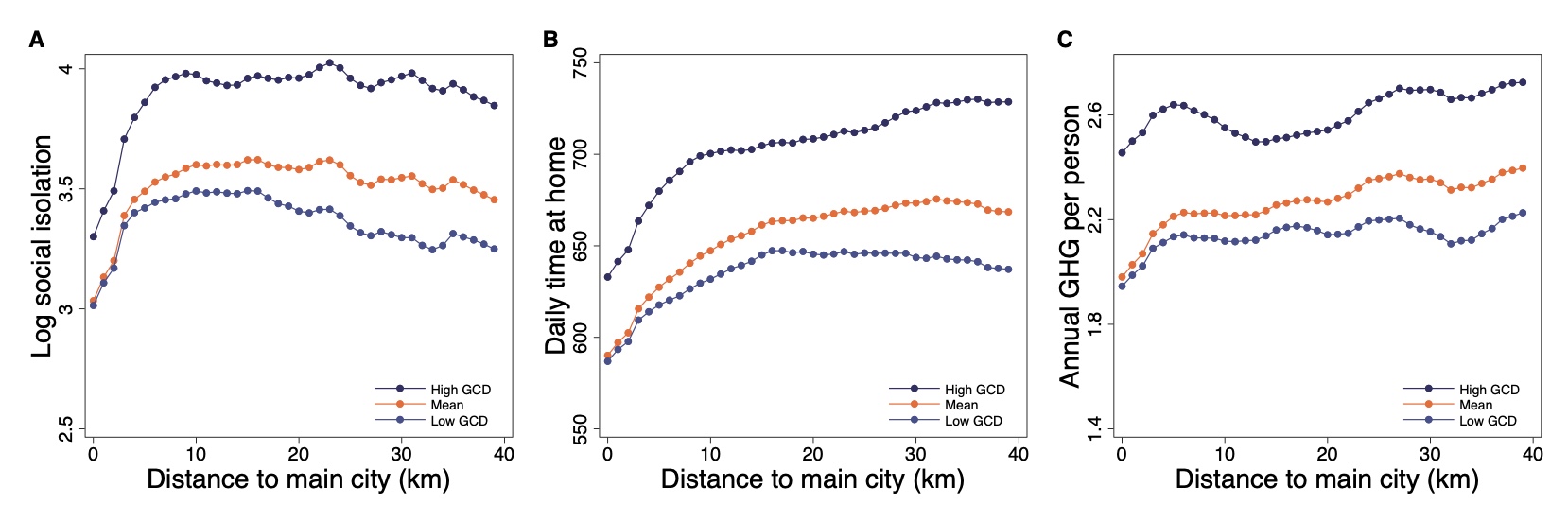}
   \caption{\small\textsc{Benchmarking the Effects of GCD versus Remoteness on Neighborhood Outcomes.} Panel A presents results for social isolation, Panel B for daily wake-time spent at home, and Panel C for annual greenhouse gas emissions per capita. The figures plot predicted outcomes for high-GCD, low-GCD, and the average neighborhood, computed for 10-km rolling windows of distance to the main city center. These are computed from regressions in each bin using the specification from Column 5 of Table~\ref{table: all discrete}. Predicted outcomes for high-GCD neighborhoods are calculated as $ \text{Predicted outcome}=\text{Mean outcome in window}+\hat{\beta}\; (1-\text{Mean garden design in window}),$  and for low GCD neighborhoods as $ \text{Predicted outcome}=\text{Mean outcome in window}-\hat{\beta}\; \text{Mean garden design in window},$ where $\hat{\beta}$ denotes the estimate of Equation \eqref{eq:ols} in that distance window.  }
	\label{fig: gdi high and low iv version}
\end{figure}

For each distance window, I use the estimated coefficients to compute expected outcome levels for high-GCD and low-GCD neighborhoods. Figure \ref{fig: gdi high and low iv version} presents these predicted outcomes: Panel A shows results for social isolation, Panel B for daily wake-time spent at home, and Panel C for annual greenhouse-gas emissions from car travel. The orange line reports the average outcome across all neighborhoods for reference.

The total environmental and social cost of GCD adoption is represented by the area between the orange line (observed outcomes) and the light-blue line (counterfactual outcomes if no neighborhoods had adopted GCD). Aggregating these differences indicates that GCD increases annual GHG emissions by 0.1 metric tons per person, raises social isolation by 13 log points (a 14\% reduction in exposure to non-residents), and adds 16 minutes of daily wake-time at home.

To benchmark these effects, I compare them to the broader costs of suburbanization. In Figure~\ref{fig: gdi high and low iv version}, this broader cost corresponds to the gap between neighborhoods near city centers and those located farther away. The total cost of suburbanization is therefore the area between the orange line (current outcomes) and the horizontal line drawn through its intercept (the outcomes if all neighborhoods were centrally located). Suburbanization as a whole is associated with an additional 0.26 metric tons of annual GHG emissions per capita, a 49-log-point increase in social isolation, and 58 minutes of time spent at home each day.

Taken together, these results imply that GCD accounts for a sizable share of the sustainability costs typically attributed to suburbanization---about 38\% (0.1 of 0.26 metric tons) of excess emissions, 27\% (13 of 49 log points) of higher social isolation, and 28\% (16.5 of 58 minutes) of additional sedentary time. The remaining costs stem primarily from remoteness and other aspects of suburban development.

\section{Concluding Remarks}\label{conclusion}
This paper examined the enduring influence of the Garden City Design (GCD) paradigm---one of the most influential planning models of the early 20th century---on contemporary social and environmental outcomes across the United States.  I introduce a new measure of GCD that captures its defining physical features and traces its historical use across the country. GCD spread rapidly after its introduction in the early 1900s, accelerated following World War II, peaked around 1990, and has declined since. Importantly, its adoption rose across all distances from city centers, consistent with historical evidence that GCD represented a nationwide design movement influencing both urban and suburban development \citet{southworth_evolving_1993, wheeler_evolution_2008}. 

The analysis shows that both neighborhood design and remoteness shape sustainability outcomes. Neighborhoods with higher GCD indices experience worse environmental and social outcomes---greater vehicle GHG emissions, higher sedentarism, and more pronounced social isolation---even when compared to neighborhoods located at similar distances from city centers. GCD-specific design accounts for 27–38\% of the negative outcomes commonly attributed to suburbanization, while macro-level factors such as distance from city centers explain the remainder. 

Together, these findings underscore that remoteness and design are complementary forces shaping the social and environmental costs of suburbanization. The empirical strategy isolates variation in design among neighborhoods with similar macro-level characteristics to show that the physical form of development---rather than distance alone---plays a significant, independent role. From a policy perspective, this means that neighborhood design is a lever that can be acted upon. Updating standards for new developments and retrofitting existing suburbs could complement broader efforts to reduce remoteness, forming an integrated strategy to mitigate the environmental and social costs of suburban sprawl. In practice, several US cities have implemented similar retrofits: Boston and San Francisco have removed highways to reconnect neighborhoods, while Portland and Washington State have improved suburban connectivity by adding local streets and pedestrian paths \citet{portland_pbot_dev_rev_manual_2022, mass_big_dig_2025, wsdot_519.1}. Whether such measures can be scaled more broadly depends on their fiscal and political feasibility, which lies beyond the scope of this paper.

The analysis also addressed several alternative explanations. In particular, it shows that the impact of GCD is not primarily driven by zoning regulations or demographic sorting but by the physical configuration of the built environment itself. This conclusion aligns with research in urban morphology showing that irregular and disconnected street networks impede navigation and access \citep{siksna_effects_1997, HAMIN2009238, stead_marshall_2002, Brown2012} and with evidence that winding, discontinuous layouts reduce walkability and limit access to amenities \citep{Kropf1996, sharifi_urban_2019}. 

More broadly, the findings show that planning paradigms leave lasting imprints on urban form and behavior. Once translated into the built environment, their spatial logic persists, shaping how people move, interact, and use resources long after the ideas themselves have faded. Understanding these enduring effects requires examining paradigms as integrated systems of design rather than as collections of isolated features. The same approach can be used to evaluate newer paradigms, such as New Urbanism, Smart Growth, or Transit-Oriented Development, and to anticipate their long-term implications for sustainability.

Future work could extend this analysis in several directions. Leveraging panel data would help disentangle behavioral changes more precisely from residential sorting. Expanding the analysis internationally could test whether similar patterns hold across urban contexts. Further refinements could improve the social isolation measure by incorporating the purpose, duration, or quality of interactions.  Another promising direction is to examine whether the effects of GCD vary across different types of places or populations---such as by region, city size, income, or race. Beyond these extensions, examining other dimensions such as mental health, infrastructure costs, or ecological impacts remains a promising avenue for research. Finally, studying interventions and retrofits aimed at improving accessibility and connectivity in existing GCD neighborhoods could offer valuable policy insights for promoting sustainable and socially cohesive urban environments.

\section*{Data Availability}

The primary dataset on mobile device locations was obtained from Safegraph. Land use data were acquired from Landgrid data provided by LOVELAND Technologies. The author does not have permission to share these two datasets. All other datasets used in this study are openly accessible. Detailed information on these datasets can be found in the Methods section.

\section*{Code Availability}

Data processing and analyses were carried out using a combination of R and Stata. The code to replicate the results in the paper can be obtained from https://github.com/ariannasm/neighborhood-design.

\section*{Competing Interests Statement}
The author declares no competing interests.

\section*{Inclusion and Ethics Statement}
The author agreed to all the manuscript contents.

{\small \renewcommand{\baselinestretch}{1} \addtolength{\parskip}{.65pt}
               \bibliographystyle{econometrica_etal}
               \ifx\undefined\BySame
\newcommand{\BySame}{\leavevmode\rule[.5ex]{3em}{.5pt}\ }
\fi
\ifx\undefined\textsc
\newcommand{\textsc}[1]{{\sc #1}}
\newcommand{\emph}[1]{{\em #1\/}}
\let\tmpsmall\small
\renewcommand{\small}{\tmpsmall\sc}
\fi

}
\clearpage

\section{Methods Section}\label{methods}

The Methods include a section describing how the sample of urban neighborhoods is defined (Section \ref{section:neigh definition}), 
a section on how the measure of GCD is computed and validated (Section \ref{section:GCD construction}), 
a section detailing the measurement of neighborhood-level outcomes (Section \ref{section:outcome construction}), a section detailing data sources for all covariates (Section \ref{section:data sources}), and a section providing details on the estimation strategies (Section \ref{section: estimation}).  

\subsection{Sample Selection and Neighborhood Definition:}\label{section:neigh definition}

Neighborhoods are defined as Census Block Groups. Specifically, I use CBGs from the 2000 US Decennial Census shapefiles. This is a convenient definition because CBGs provide consistent national coverage, feature standardized boundaries, and have a granular population size (between 800 and 3,000 residents). 

I then restrict the analysis to CBGs in urban areas, excluding rural regions. Urban areas are delineated using building density raster data from \citet{leyk_2018}, available through the HISDAC-US database (\url{https://dataverse.harvard.edu/dataverse/hisdacus}). Building density is measured using 250m $\times$ 250m raster grid cells covering the entire US, aggregated at 5-year intervals from 1800 to 2015. Each pixel records the total number of structures within it, based on underlying data from Zillow Transaction and Assessment Data (ZTRAX). 

\begin{figure}[!ht]
	\centering
	\figuretitle{}
	\includegraphics[width=.7\linewidth]{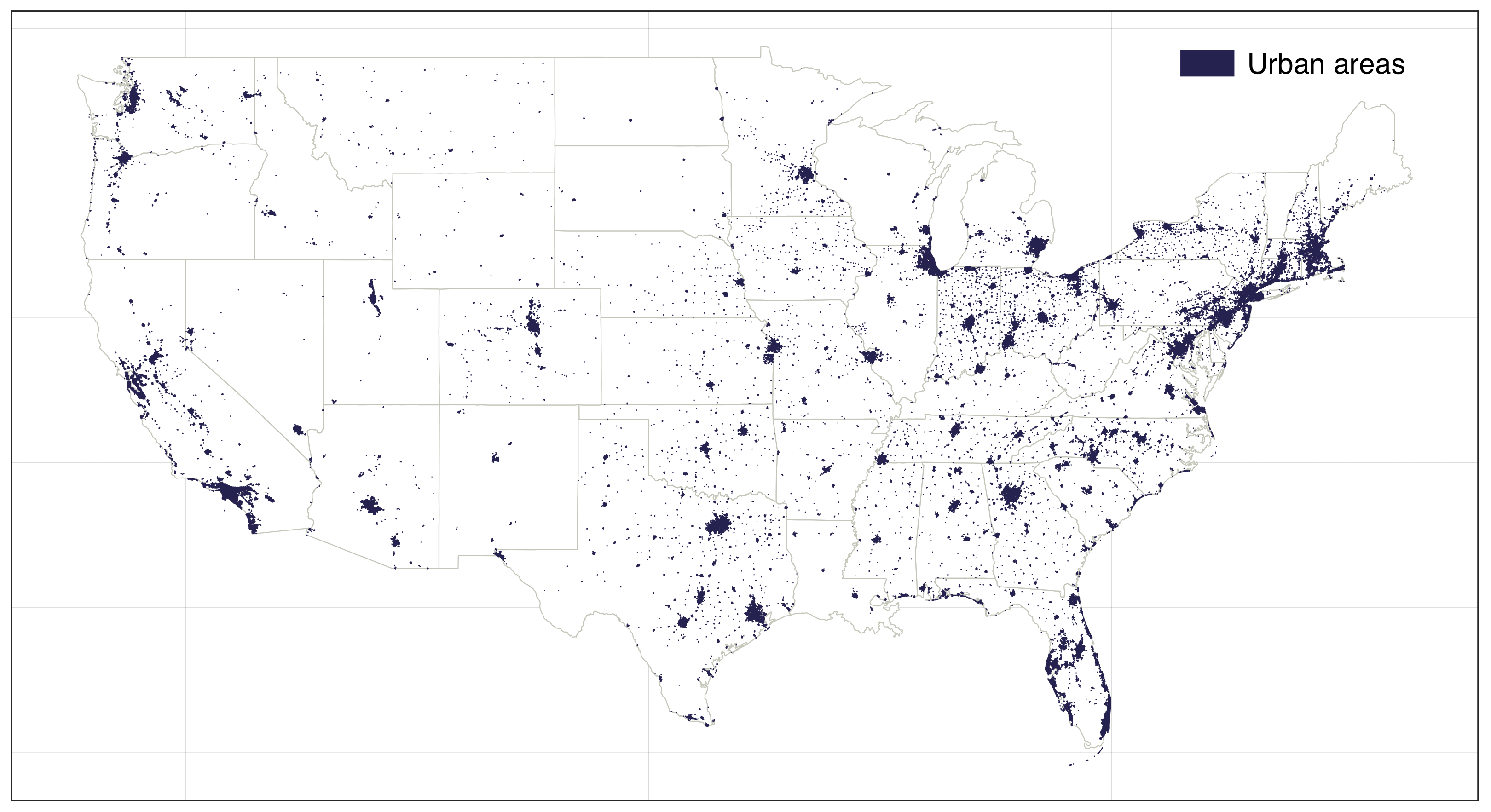}
   \caption{\small\textsc{Map of Urban Areas.} The map shows the location of the urban areas defined using the 250x250 raster data by \citep{leyk_2018}. 22,494 urban areas and 60,421 neighborhoods within these areas were identified.}
	\label{fig: urban area map}
\end{figure}

An urban area is defined as a contiguous set of 250m × 250m grid cells, each of which individually exceeds a 50\% building density threshold in 2000. Applying this criterion yields 22,494 distinct urban areas, shown in Appendix Figure~\ref{fig: urban area map}. CBGs are spatially overlaid onto these urban boundaries, and only those block groups with at least 50\% of their area overlapping these boundaries are retained in the final sample.  The final dataset consists of 60,421 urban neighborhoods.

All results are shown using this CBG-based definition of neighborhoods. For completeness, Table \ref{table: tract level} shows that the results are robust to using Census tracts as an alternative (and more aggregated) definition.

\subsection{Garden City Design (GCD) Measure:}\label{section:GCD construction}

My GCD measure uses data from OpenStreetMap (available at \url{https://openstreetmap.org}) for 2021. These data include the location and shape of all street segments in the US. 

To quantify adherence to Garden City design principles, I construct a composite measure consisting of four components:

\begin{enumerate}
    \item Long-winded Streets: Historically, Garden City designs favored winding streets that followed natural terrain, providing varied vistas and dynamic visual orientations \citep{cleveland_1871, creese_1966}. To quantify this feature, I calculate the ratio of the actual length of each street segment to the straight-line (Euclidean) distance between its endpoints. These ratios are averaged across all streets within each neighborhood.
    
    \item Enclosed Streets: Garden City neighborhoods aimed to create cohesive, secluded residential enclaves distinct from surrounding areas, frequently employing cul-de-sacs and dead-end streets \citep{stern_paradise_2013}. I quantify enclosed streets by measuring the proportion of streets in each neighborhood that terminate internally---that is, streets having only one endpoint connecting to the broader network. 
    
    \item Block Organicity: Garden City neighborhoods typically featured varied, organically-shaped blocks rather than uniform rectangular grids. These irregular blocks accommodated open spaces and diverse land uses, such as parks and playgrounds. To quantify block organicity, I measure the proportion of block angles within each neighborhood that significantly deviate from the standard rectangular shape (90° angles). Specifically, the calculation involves the following steps. First, using the OpenStreetMap street network data, I define blocks by polygons completely enclosed by the main street segments, keeping only segments classified as primary, secondary, tertiary, residential, or trunk roads. Minor streets, such as pedestrian pathways, footpaths, alleys, and service roads, are  excluded to prevent artificially small blocks. 
    
    Next, for each block polygon, I calculate the internal angles formed by its vertices. Because of the way streets are drawn in OpenStreetMap, right angles are not always recorded as exactly 90 degrees, and straight lines are often split into two segments with angles close to 180 degrees. To address this, I allow a tolerance of 10 degrees around 90 (treating angles between 80° and 100° as rectangular) and disregard angles close to 180° (between 165° and 195°) to avoid double-counting straight lines. Finally, for each neighborhood, I calculate the proportion of block angles that fall outside these ranges. Higher proportions indicate stronger adherence to the organic, irregular block layouts characteristic of Garden City designs.
    
    \item Hierarchical Street Network: Garden city planning introduced hierarchical road networks that clearly differentiated major from minor roads. Minor roads frequently curved gently into larger roads without disrupting their continuity, establishing a distinct flow and spatial hierarchy \citep{robinson_1926}. To quantify this feature, I calculate the proportion of three-way intersections within each neighborhood---points at which minor roads merge into major ones without intersecting.
\end{enumerate}

These four standardized components are averaged to create the composite GCD measure, as summarized in the following equation:
\begin{align*}
    \begin{array}{l}
    \textrm{Garden City} \\
    \textrm{design}
    \end{array}=\begin{array}{l}\textrm{Long-winded}\\ \textrm{streets}\end{array}+ \begin{array}{l}\textrm{Enclosed}\\\textrm{streets}\end{array} + \begin{array}{l}\textrm{Block}\\\textrm{organicity}\end{array}+ \begin{array}{l}\textrm{Street}\\\textrm{hierarchy}\end{array}.
\end{align*}

In total, each component contributes a similar share to the overall variation in the index: organic blocks (26\%), hierarchical streets (27\%), enclosed streets (22\%), and curvilinear layouts (25\%). This balance ensures that the composite score reflects the combined presence of all four design traits, rather than being driven by any single characteristic.
Figure \ref{fig: components} in the Appendix illustrates the change in these components over time and across space.  

\paragraph{GCD Measure Validation:}
I validate the GCD measure by evaluating its ability to identify neighborhoods historically classified as Garden City developments, using classifications from prior literature:

\begin{itemize}
\item \citet{talen_2022} compiled a dataset of several planned communities across the US, manually classified into distinct typologies.
In total, her data includes 46 neighborhoods labeled as either ``garden city,'' ``garden village,'' or ``resort garden suburb,'' all conforming to GCD ideas and used in the validation.

\item \citet{wheeler_evolution_2008} manually classified 39,985 neighborhoods into 25 distinct typologies based on visual inspection of design features via Google Earth imagery. These typologies capture characteristics such as street layout, lot size, building footprint, and land-use mix across six major US metropolitan areas. I exclude typologies labeled ``commercial strip,'' campus,'' and ``civic,'' as these categories do not overlap with my urban sample. This results in a validation sample comprising 8,795 neighborhoods belonging to different typologies. Of these, several typologies are related to GCD, including ``loops and lollipops,'' ``garden apartments,'' and ``garden suburb.'' The remaining typologies are related to other paradigms. 

\item \citet{SalazarMiranda2021} digitized a set of 55 neighborhoods developed by the United States Housing Corporation (USHC) in 1919, historically recognized as early exemplars of Garden City planning in the US.
\end{itemize}

\begin{figure}[!ht]
	\centering
	\figuretitle{}
	\includegraphics[width=.7\linewidth]{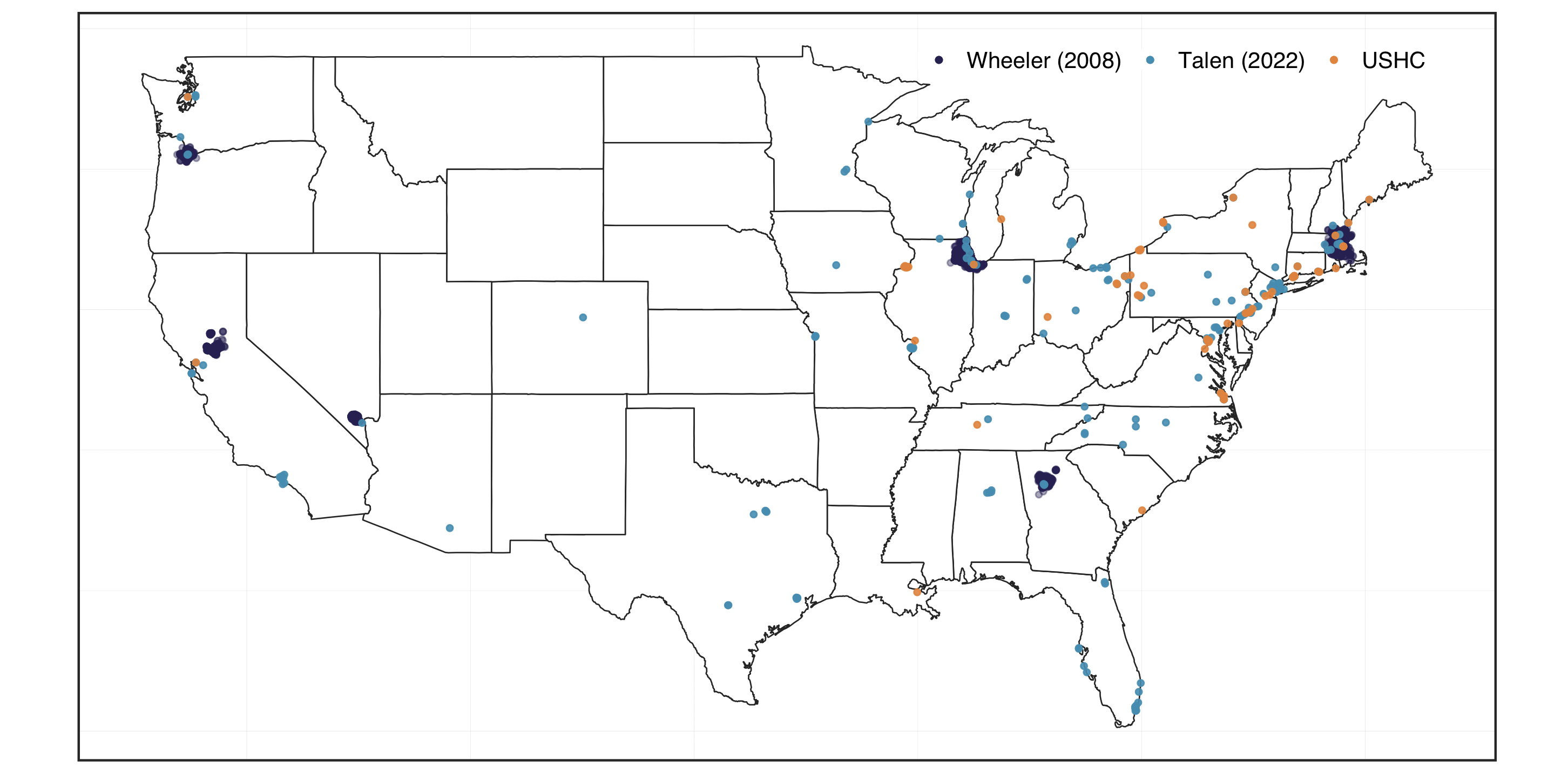}
 \caption{\small\textsc{Geographical Distribution of Neighborhoods for GCD Validation.} Location of iconic Garden City neighborhoods using data from \citet{wheeler_evolution_2008}, \citet{talen_2022}, and \citet{SalazarMiranda2021}.}
	\label{fig: validation map}
\end{figure}

 Figure \ref{fig: validation map} shows the location of neighborhoods studied by \citet{talen_2022}, \citet{wheeler_evolution_2008}, and \citet{SalazarMiranda2021}.
To link these historical classifications to neighborhoods in my dataset, I match them to the CBG containing their centroids.

\begin{figure}[!ht]
	\centering
    \includegraphics[width=0.9\linewidth]{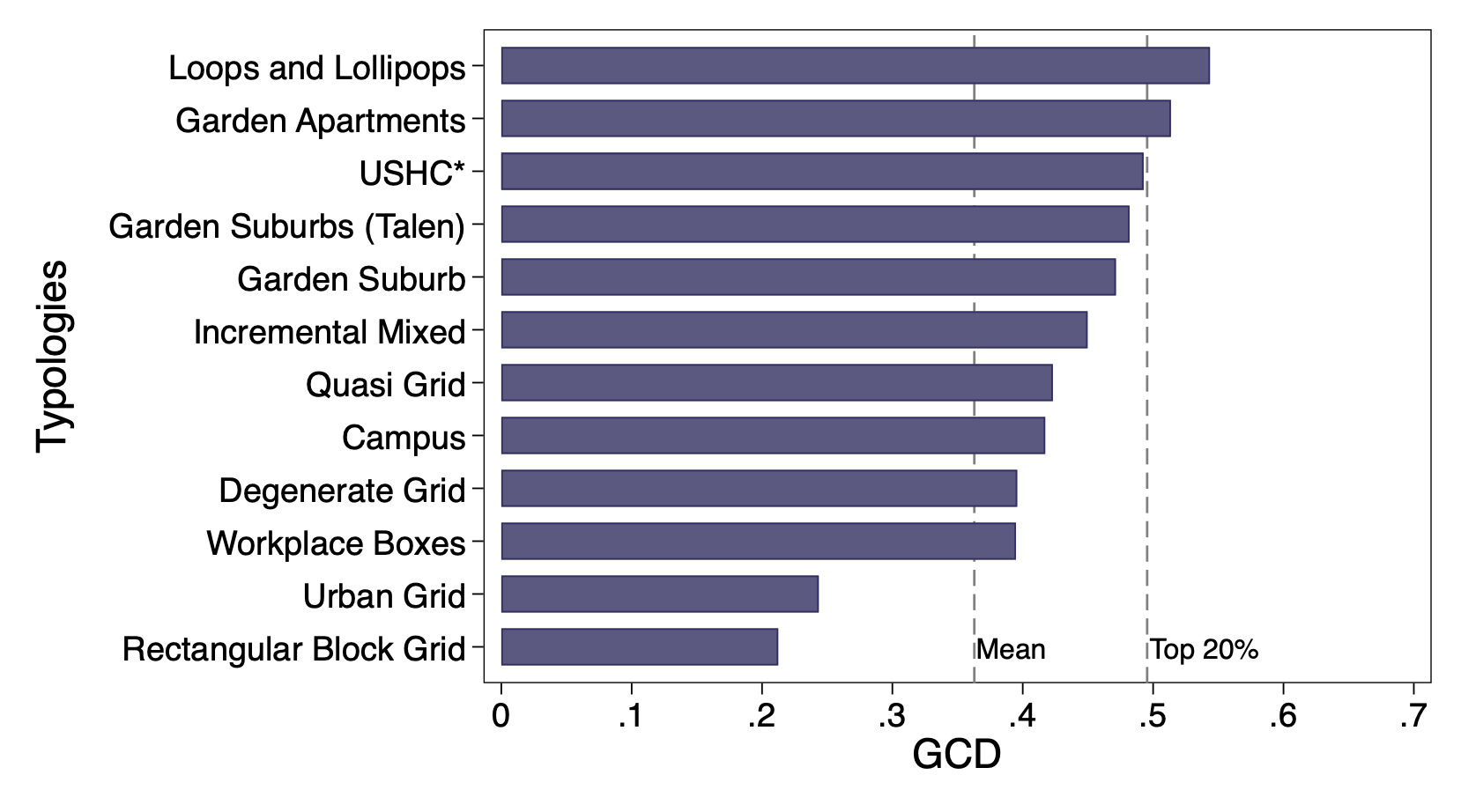}
	\caption{\small \textsc{Validation of GCD Measure Using Existing Neighborhood Classifications.} Average GCD index shown for neighborhood typologies identified by \citet{wheeler_evolution_2008}, several GCD typologies identified by \citet{talen_2022} (labeled as ``Garden Suburbs''), and neighborhoods developed by the USHC digitized by \citet{SalazarMiranda2021}. Reference lines represent the overall mean and 80th percentile of the GCD index in my sample of urban US neighborhoods. }
	\label{fig: gdi validation}
\end{figure}

Figure \ref{fig: gdi validation} compares the average GCD index of neighborhoods identified by these three external classifications to the overall distribution of the GCD measure in my sample. Typologies explicitly aligned with Garden City principles, such as ``loops and lollipops,'' ``garden apartments,'' and ``garden suburbs,'' as well as the GCD neighborhoods identified by \citet{talen_2022} and the USCH neighborhoods studied by \citet{SalazarMiranda2021} consistently rank above or near the top 20\% of the GCD distribution. Conversely, typologies characterized by orthogonal layouts and limited adherence to Garden City design, including grids, quasi-grids, and streetcar suburbs, fall in the lower end of the distribution.

\subsection{Outcome Variables:\label{section:outcome construction}}

This section provides data sources and describes how the outcome variables are constructed.

\paragraph{Social Isolation:} This measure is constructed using anonymized smartphone location data from SafeGraph for 2019 (see \url{https://safegraph.com/}). SafeGraph tracks daily movements of smartphone users. Their sample is broadly representative of the US population, providing broad demographic and geographic coverage with small sampling biases in age, gender, and income \citep{Li2023}. 

Safegraph identifies the primary residential location (``home’’) of each user as the census block group where they spent most nighttime hours over a 6-week observation period. It then provides a series of statistics, including the number of visits by residents in each neighborhood to points of interest (POIs) in their own and other neighborhoods---including coordinates for schools, parks, retail establishments, and other amenities. Safegraph provides the location of these POIs, allowing me to compute total visits from residents of a neighborhood to POIs in their neighborhood, as well as visits from non-residents to the same set of POIs. 

Using these data, I measure  neighborhood-level social isolation as the inverse of potential interaction opportunities. Formally, these opportunities are measured using a standard exposure index:
\begin{align*}
    \textrm{Exposure}_{i} = \frac{\textrm{Resident trips}_{i}}{\textrm{Residents}_{i}} \; \frac{\textrm{Non-resident trips}_{i}}{\textrm{Resident and non-resident visitors}_{i}}.
\end{align*}
The first term gives the frequency with which residents visit local POIs. This is computed as the number of visits made by residents of neighborhood $i$ to POIs inside this neighborhood (obtained from SafeGraph), divided by the neighborhood population (obtained from the US Census). 

The second term gives the share of non-residents among visitors to local POIs. This is computed as the number of visits made by non-residents to POIs inside neighborhood $i$, divided by the number of visitors (both residents and non-residents). Both the numerator and denominator come from SafeGraph.

The above measure builds on recent literature on experienced segregation \citep[see, for example,][]{athey_experienced_2021, moro2021, abbiasov2024}, which quantifies social interactions based on spatial patterns of visitation between distinct demographic groups using exposure-style measures like the one introduced above. 

Social isolation is then computed as:
\begin{align*}
    \textrm{Social Isolation}_{i} = -\log\big(\textrm{Exposure}_{i}\big).
\end{align*}
Higher values of this measure indicate fewer interaction opportunities and thus greater social isolation. The negative sign ensures that higher numerical values correspond directly to increased isolation.

\paragraph{Sedentarism:} This measure also uses data from SafeGraph, which reports the average daily wake-time users spent at home in 2019, excluding typical sleeping hours.  As explained above, 
SafeGraph identifies home locations by 
average time spent during sleep hours. The proxy for sedentarism averages these daily waking-time minutes spent at home for all residents.

\paragraph{Greenhouse gas emissions:}
Neighborhood-level estimates of annual greenhouse gas emissions per capita associated with car travel are obtained from the Environmental Protection Agency's (EPA) Smart Location Database for 2017 (\url{https://www.epa.gov/smartgrowth/smart-location-database-technical-documentation-and-user-guide}).  Estimates are provided at the census block-group level and incorporate commuting vehicle miles traveled (VMT) and non-commute ones. Commute distances are calculated based on workplace locations and commuting patterns from the National Household Travel Survey (NHTS). The EPA also provides an estimate of average GHG emissions per VMT, used to compute total emissions associated with both types of trips. A detailed technical description of the EPA procedures is available at \url{https://www.epa.gov/smartgrowth/smart-location-mapping#SLD}.

\subsection{Other Covariates Used in the Analysis\label{section:data sources}}

This section describes additional data sources and variables used in the study. 

\begin{figure}[!ht]
	\centering
	\includegraphics[width=0.5\linewidth]{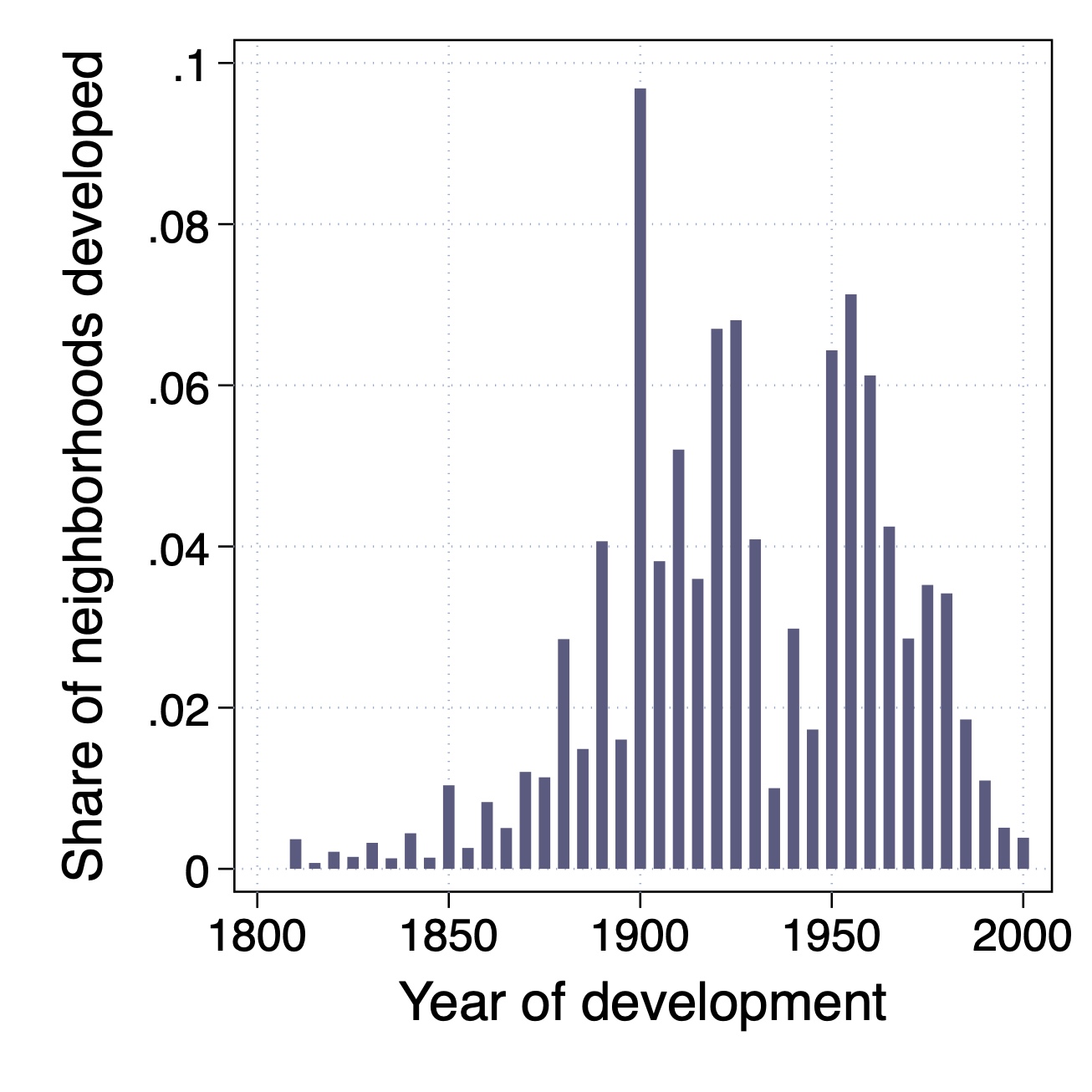}
	   \caption{\small\textsc{Neighborhood development year.} The figure plots the share of neighborhoods developed each year. This is defined as the year when the largest number of pixels in the neighborhood were first developed (i.e., the modal development year), using data from \citet{leyk_2018}.}
	\label{fig: neigh developed by year}
\end{figure}

\paragraph{Development Year:}
The year of development for each neighborhood was determined using the historical component of the data from \citet{leyk_2018}. A neighborhood's development year is defined as the modal date of first-time development among all its pixels---that is, the year when the largest number of pixels within the neighborhood were first developed. Figure~\ref{fig: neigh developed by year} shows the percentage of neighborhoods developed each year based on this measure.

\paragraph{Distance to City Centers:} This measure relies on data from the National Atlas of the United States (2004), provided by the Stanford Geospatial Center (\url{http://purl.stanford.edu/js689mk1912}). This dataset provides geographic coordinates and population information for all incorporated cities and towns in the US. Using these data, the distance to the main city center is computed as the distance from the centroid of a neighborhood to the centroid of the most biggest city in the same urban area.

\paragraph{Geographic Characteristics:} 
Geographic and ecological characteristics are obtained from multiple sources. Elevation and slope data are from the US Geological Survey's 1/3 arc-second (approximately 10-meter resolution) Digital Elevation Model (DEM) dataset, available through the USGS National Map (see \url{https://apps.nationalmap.gov/downloader}). Ecological classification data are from the EPA's Level I Ecoregions dataset (see \url{https://www.epa.gov/eco-research/ecoregions}), which divides the US territory into nine major ecological areas based on broad environmental attributes such as vegetation, soil, climate, and topography. Neighborhood centroid latitude and longitude coordinates are computed directly from Census shapefiles.

\paragraph{Demographic Variables and Contemporary Population Density:} The 
demographic data used in Appendix Table~\ref{table: ols discrete robustness}
are from the 2000 US Decennial Census, obtained via the NHGIS database (see \url{https://www.nhgis.org/}). Variables include median age, per capita income, percentage of residents with some college education, percentage married, and population density (residents per square kilometer). 

\paragraph{Historical density:} The historical density control in Table \ref{table: ols discrete robustness} is also obtained using the data from \citet{leyk_2018}. 

\paragraph{Tract-Level Work-From-Home Rates:}
Census tract–level work-from-home shares are derived from the American Community Survey (ACS) 5-year estimates for 2015–2019 (see \url{https://data.census.gov}). The ACS asks employed individuals about their usual means of transportation to work, including an option for ``work at home.'' For each tract, the WFH share is calculated as the proportion of workers who report that they usually work from home, relative to the total number of workers.  The resulting variable is included as a control in a robustness check reported in Table~\ref{table: ols discrete robustness}.

\paragraph{Zoning Data:} The zoning data used in Figure~\ref{fig: zoning gdi time} and Table~\ref{table: zoning} is from the 2020 Landgrid parcel database from LOVELAND Technologies (\url{https://app.regrid.com/store}). Parcels are classified according to their primary economic or functional use following the Land-Based Classification Standards (LBCS). A parcel is designated as residential if classified as as the LBCS as ``residential,'' ``private households,'' or ``accommodations'' (see \url{https://www.planning.org/lbcs/standards/function/} for detailed categories). From these data, I measure the share of plots in each neighborhood zoned as residential.
 
\paragraph{County-Level Migration Rates:}
County-level migration rates used in Table~\ref{table: low migration} are from the IRS (see \url{https://www.irs.gov/statistics/soi-tax-stats-migration-data}). These records measure annual inflows and outflows of households for each county for 1999–2000. Migration rates are computed as the sum of inflows and outflows divided by the county population. 

\subsection{Estimation Details and Identification Strategy:}\label{section: estimation}

\paragraph{OLS estimates:} The estimated OLS equation is provided in the main text (equation \ref{eq:ols}). OLS estimates identify the causal effect of GCD on outcomes provided that  
\begin{align}
    \label{eq:cia}
    \mathbb{E}[\epsilon_{ims}| \textrm{Garden City design}_{ims}, d(i), m,s, X_{ims}]=0.
\end{align}
This assumption requires that there are no unobserved confounders affecting both GCD and outcomes.

\paragraph{Propensity score matching estimates:} average treatment effect of GCD are computed using the inverse propensity score reweighing scheme of \citet{Hirano2003}:
\begin{align*}
    \hat{\tau}=\sum_{i \in \text{high GCD}} \frac{Y_i}{\hat{P}(X_i)}-\sum_{i \in \text{low GCD}} \frac{Y_i}{1-\hat{P}(X_i)}.
\end{align*} 
Here, $\hat{P}(X_i)$ the estimated propensity score as a function of the covariates from Column 3 of Table \ref{table: all discrete} and $Y_i$ is the outcome of interest. 

\begin{figure}[!ht]
	\centering
	\figuretitle{}
	\includegraphics[width=0.5\linewidth]{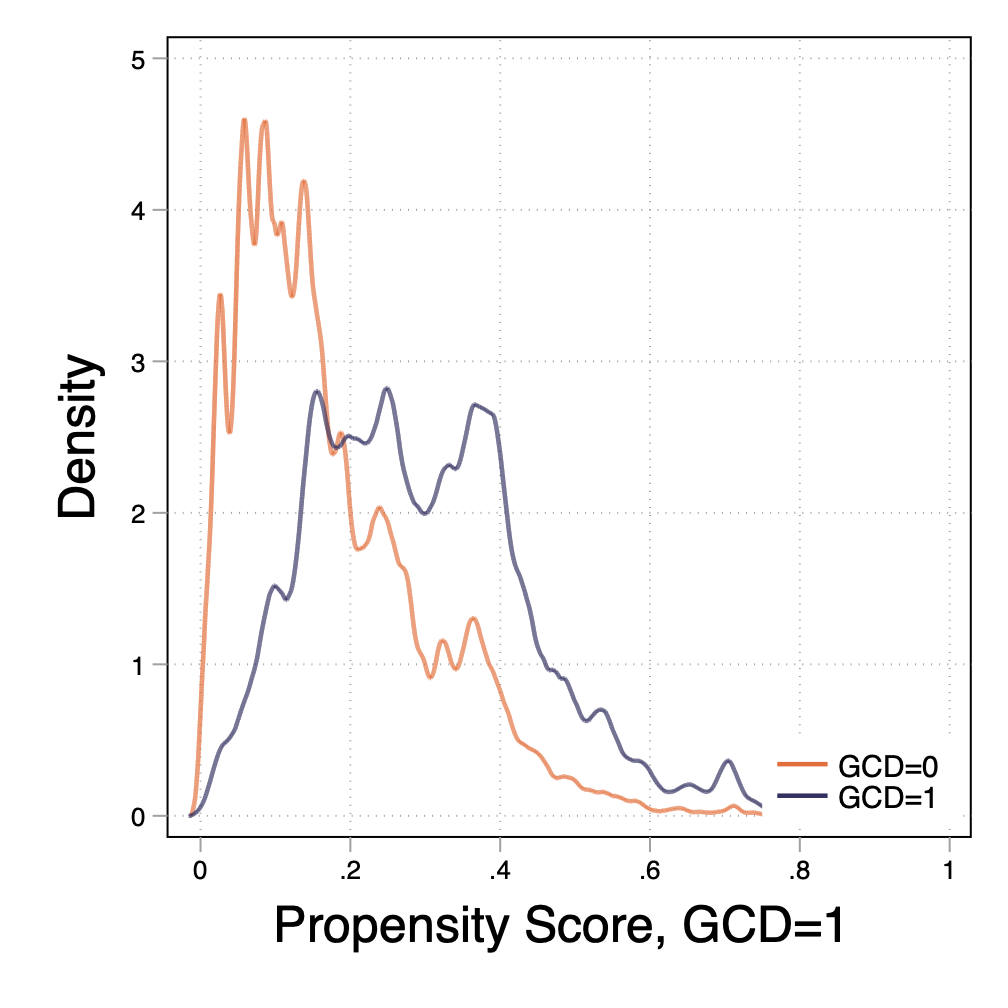}
	\caption{\small \textsc{Propensity score distributions.} The figure shows the propensity score distribution for high- and low-GCD neighborhoods. The propensity score is estimated using the covariates from Column 3 of Table \ref{table: all discrete}.}
	\label{fig: propensity score}
\end{figure}

A requisite assumption is that there is overlap in the distribution of propensity scores among treated and control neighborhoods. Figure \ref{fig: propensity score} inspects this assumption visually and shows that these distributions overlap in the (0,1) range.

\paragraph{IV estimates:} The estimated IV equation is provided in the main text (equation \ref{eq:ols}). The IV estimates identify the causal effect of GCD on outcomes provided that  \begin{align}
    \label{eq:cia}
    \mathbb{E}[\epsilon_{ims}| \overline{ \textrm{GCD}}(t_i) , d(i), m,s, X_{ims}]=0,
\end{align}
This assumes design waves only affect outcomes through GCD, and not other unobserved practices that could be in the error term $\epsilon_{ims}$ (i.e., the exclusion restriction). 

\paragraph{Age effects:} In this empirical setting, age effects could confound the IV estimates if older neighborhoods---being part of earlier cohorts---have greater social capital, leading to better contemporary outcomes. For instance, older neighborhoods might have had ample time to cultivate social relations, enhancing public space management. 

To address this, I exploit the fact that before 1875, there were no significant differences in neighborhood design. As shown in Figure \ref{fig: gdi time}, the use of GCD from 1810--1870 remained flat. In fact, historical accounts suggest that this was an era marked by the use of grid and other planning paradigms. Using this sample, I estimate the following variant of Equation \eqref{eq:ols} via OLS
\begin{align*}
    \textrm{Outcome}_{ims} = &
    \beta_a\;\textrm{Age}_{ims} + F(d(i)) + \alpha_{m} + \gamma_{s} + \theta\; X_{ims} + \epsilon_{ims}
    \addtag\label{eq:age effects}
\end{align*}
This equation explains the outcome of a neighborhood as a function of its age and the covariates used in columns 1 and 2 of Table \ref{table: all discrete}. The regression excludes the measure of GCD from the right-hand side because this explanatory variable is unchanged during this period. 

The coefficient $\beta_a$ captures the effect of age, which proxies for accumulating neighborhood capital with age. Columns 1 and 2 of Table \ref{table: age effects} report the estimates of Equation \eqref{eq:age effects} for the main outcomes of interest (in different panels). The estimated age effects are all small and precisely estimated, which suggests that this is not a big source of bias in my IV estimates. 

To account for the role of age effect, I re-estimate a variant of Equation \eqref{eq:ols} via IV:
\begin{align*}
    \textrm{Outcome}_{ims} - \widehat\beta_a\;\textrm{Age}_{ims} = &
    \beta\;\textrm{GCD}_{ims} + F(d(i)) + \alpha_{m} + \gamma_{s} + \theta\; X_{ims} + \epsilon_{ims}.
    \addtag\label{eq:age adjusted ols}
\end{align*}
This equation now adjusts the outcome variable by subtracting the estimated contribution of age based on the 1810--1870 sample. It then instruments for GCD using neighborhoods' historical cohorts. Columns 3-4 of Table \ref{table: age effects} report the IV estimates, and Columns 5–6 report OLS estimates of Equation \eqref{eq:age adjusted ols} for comparison. Odd columns control for distance to city center. Even columns add all geographic and location controls. The estimates adjusting for age effects show that GCD continues to be significantly associated with worst social and environmental outcomes, and that this is not a mechanical artifact due to the fact that GCD neighborhoods are ``younger.''

\clearpage
\section*{Supplementary Information}\label{appendix}
\renewcommand{\thefigure}{A\arabic{figure}}
\setcounter{figure}{0}
\renewcommand{\thetable}{A\arabic{table}}
\setcounter{table}{0}

\begin{figure}[!ht]
    \centering
    
    \begin{subfigure}{\linewidth}
        \centering
        \includegraphics[width=.6\linewidth]{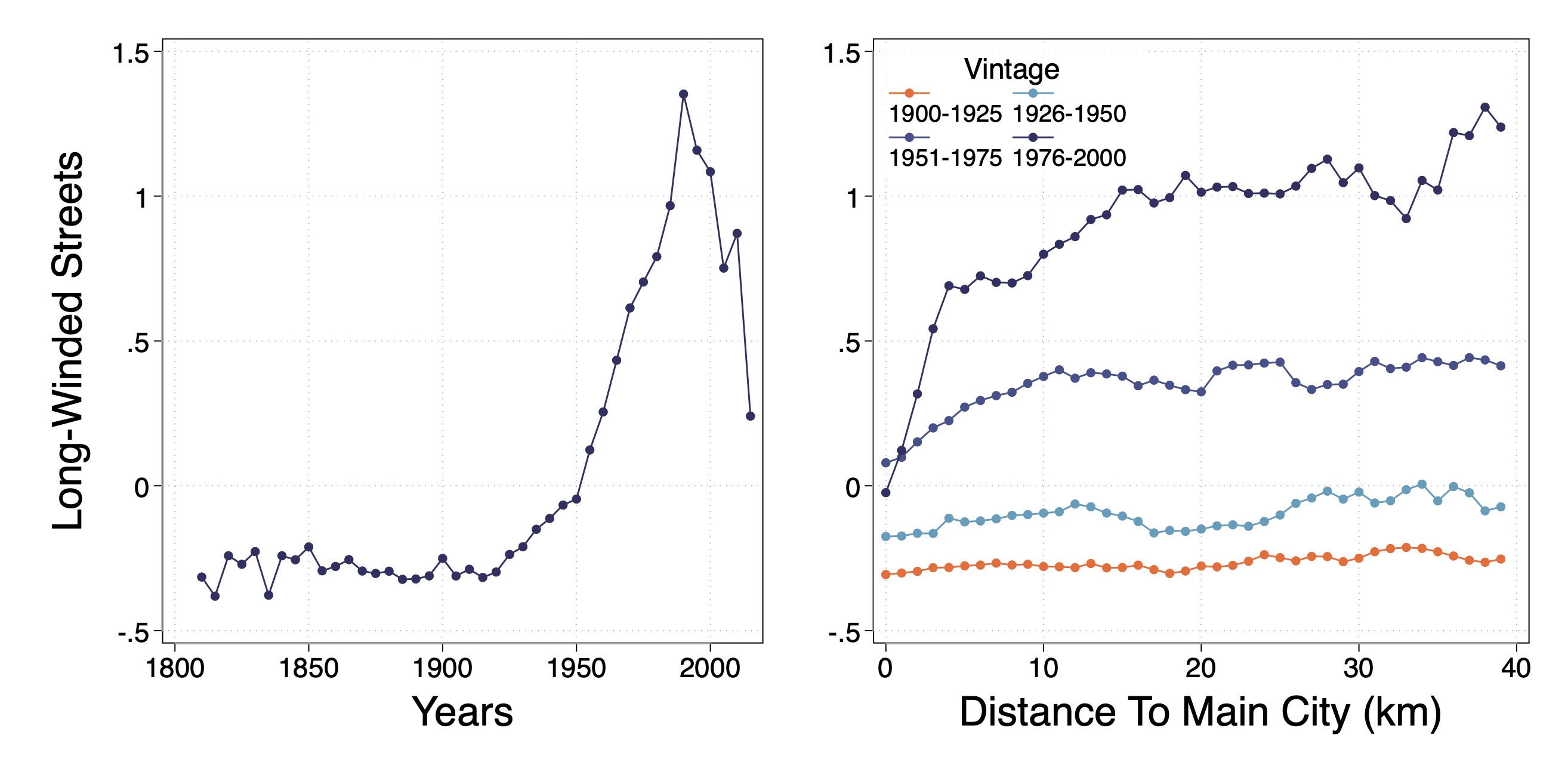}
    \end{subfigure}%
    \begin{subfigure}{\linewidth}
        \centering
        \includegraphics[width=.6\linewidth]{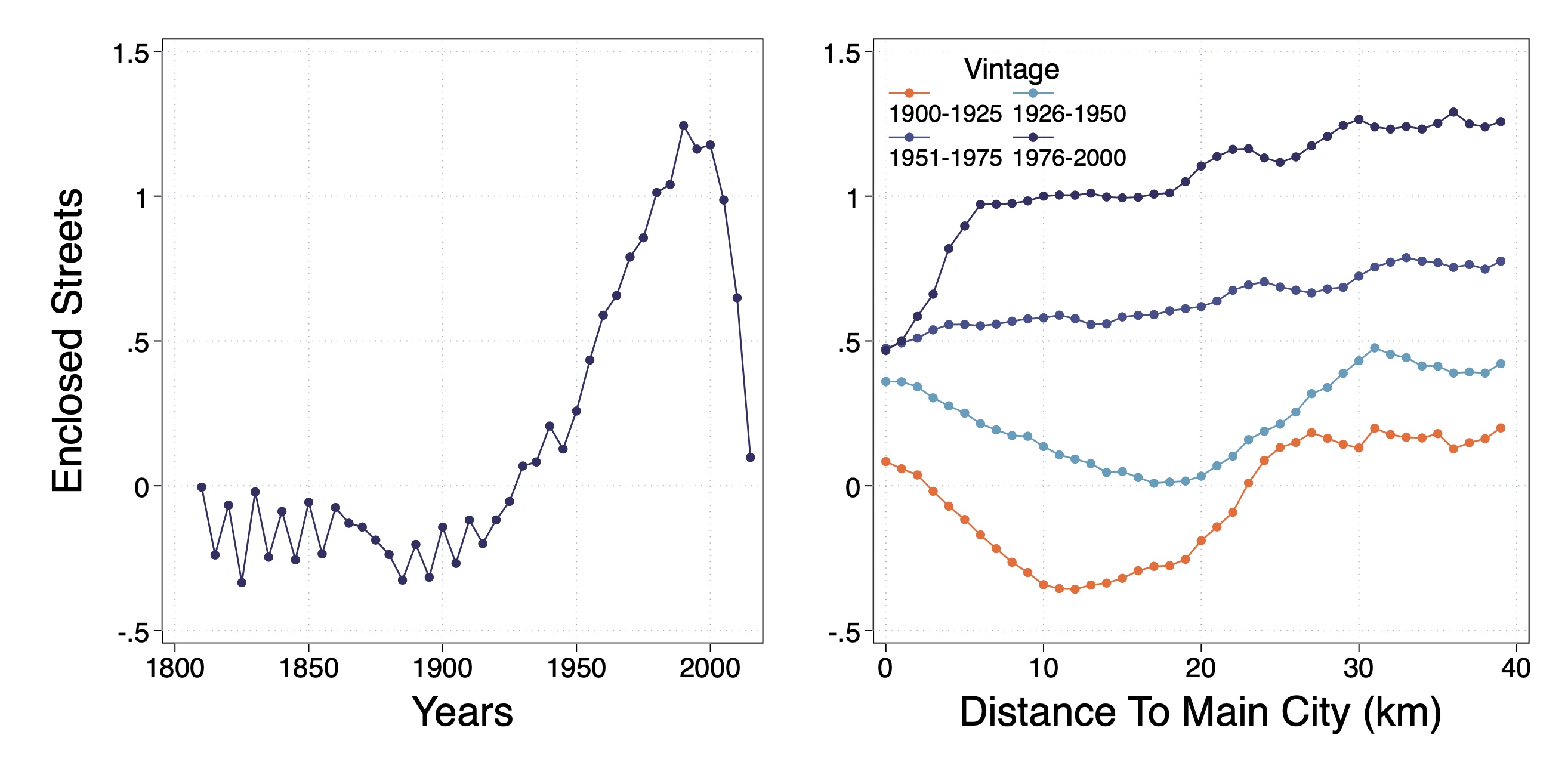}
    \end{subfigure}%
    \begin{subfigure}{\linewidth}
        \centering
        \includegraphics[width=.6\linewidth]{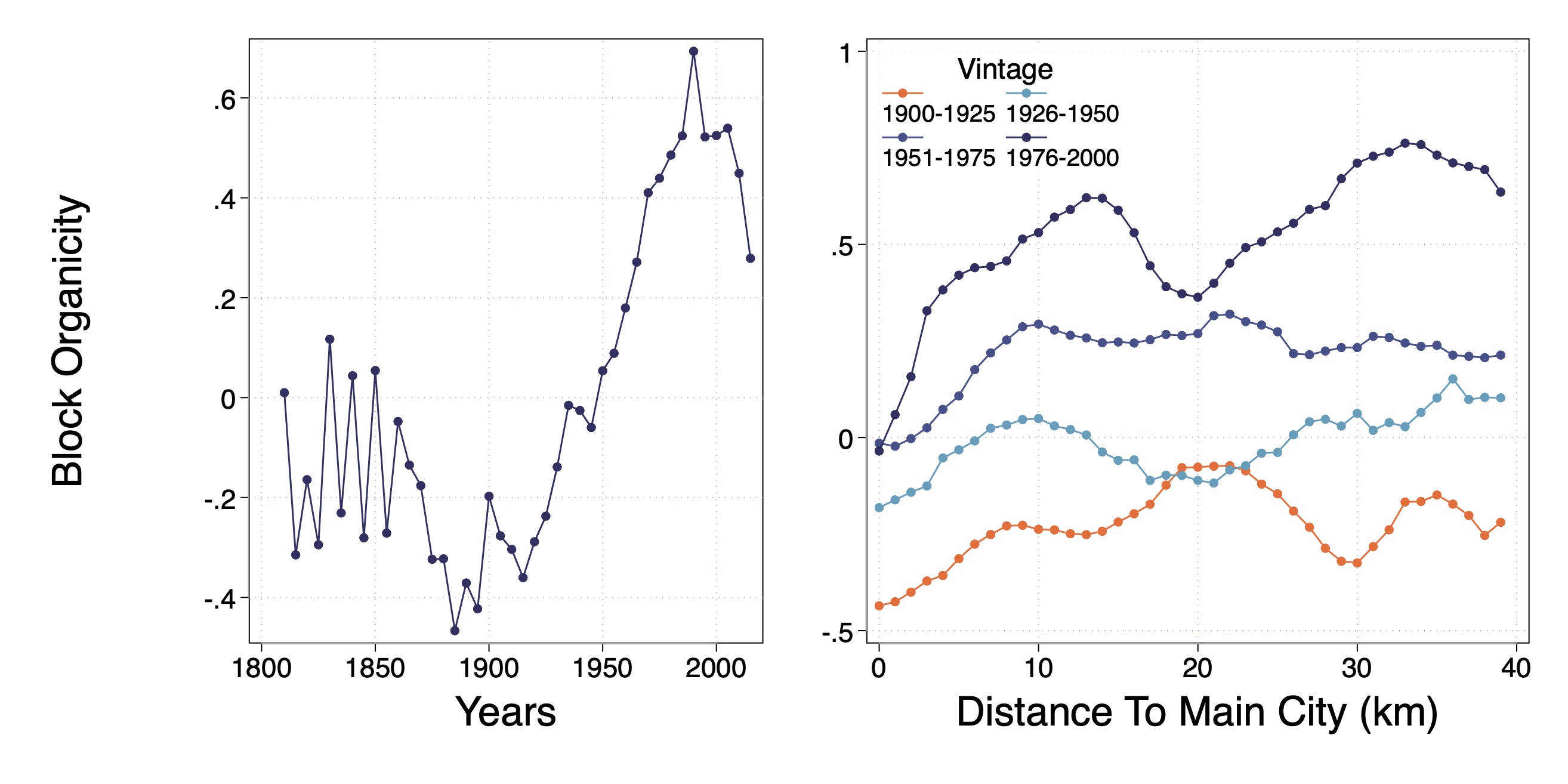}
    \end{subfigure}%
    \begin{subfigure}{\linewidth}
        \centering
        \includegraphics[width=.6\linewidth]{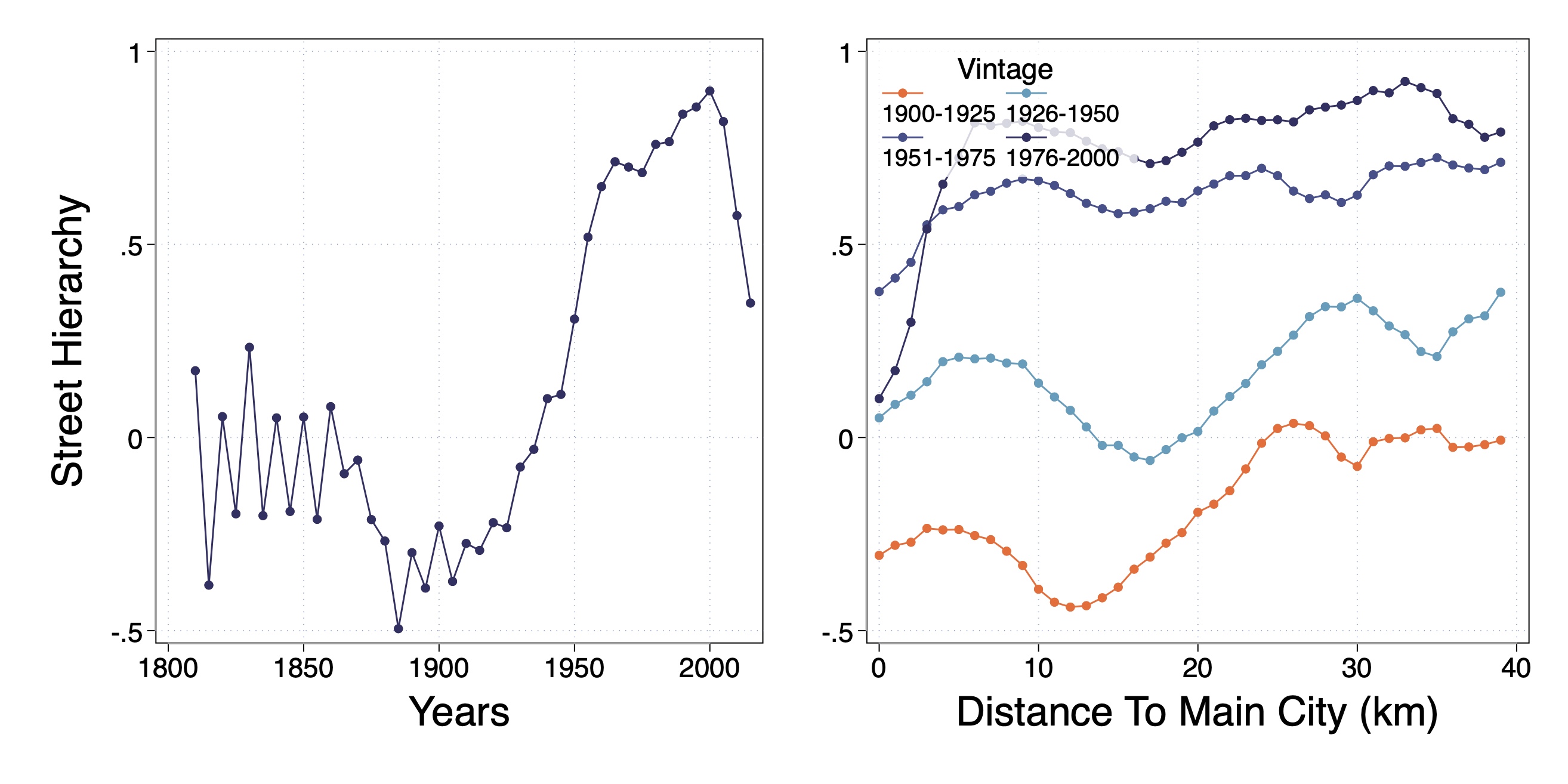}
    \end{subfigure}    \caption{\small\textsc{Spatial and Temporal Evolution of GCD Components.} This figure is analogous to Figure \ref{fig: gdi time}, focusing now on the components of the GCD measure: long-winded streets, enclosed streets, block organicity, and street hierarchy. The first panel in each row traces the temporal evolution of each component across five-year intervals from 1800 to 2015. The second panel shows the relationship between each component with distance from the primary city center across distinct periods: pre-1925, 1926-1950, 1951-1975, and 1976-2000. } 
    \label{fig: components}
\end{figure}

\begin{figure}[!ht]
	\centering
	\figuretitle{}
	\includegraphics[width=0.7\linewidth]{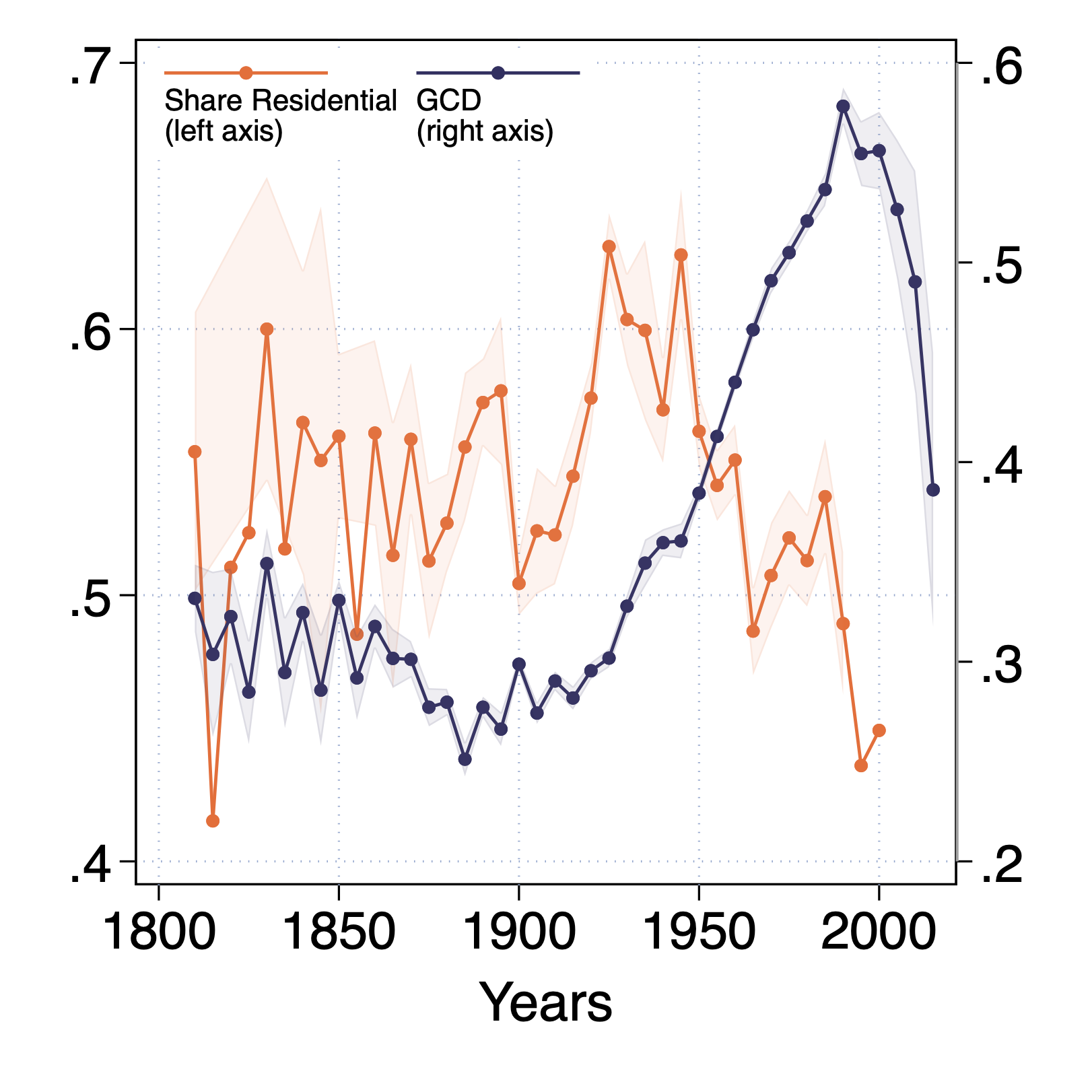}
	   \caption{\small\textsc{Evolution of GCD and Residential Zoning.} The figure plots the time trajectories of GCD (shown in blue, right axis) and the prevalence of residential zoning (shown in orange, left axis) for neighborhoods developed at different points in time. Data was aggregated over 5-year intervals from 1810 to 2015. The error bands represent 95\% confidence intervals.}
	\label{fig: zoning gdi time}
\end{figure} 

\clearpage

\begin{table}[!ht]
	\centering
	\caption{\sc{OLS and IV Estimates using the continuous measure of Garden City Design as explanatory variable.}}
    \label{table: all continuous}
	\resizebox{1\textwidth}{!}{\begin{tabular}{L{6cm}C{3cm}C{3cm}C{3cm}C{3cm}C{3cm}}\toprule\toprule
			\vspace{0.5cm}
			&(I) &(II)&(III)&(IV)\\
			\cmidrule(r){2-3} \cmidrule(r){4-5}  \\
			&\multicolumn{2}{c}{OLS Estimates} & 	\multicolumn{2}{c}{IV Estimates} \\
			\vspace{0.5cm}
			&\multicolumn{4}{c}{\sc{Panel I. Dependent Variable:  Log Social isolation}}\\\cmidrule(r){2-5}	
			GCD Index   &       0.442$^{***}$&       0.409$^{***}$&       0.791$^{***}$&       0.940$^{***}$\\
            &     (0.100)        &     (0.084)        &     (0.164)        &     (0.154)        \\
Observations&       45226        &       45226        &       45226        &       45226        \\
R-squared   &      514.00        &      505.00        &      514.00        &      505.00        \\
F-Stat      &        0.01        &        0.05        &        0.01        &        0.02        \\
widstat     &                    &                    &      385.84        &     1423.02        \\

			\vspace{0.2cm}
			&\multicolumn{4}{c}{\sc{Panel II. Dependent Variable:  Daily Time at Home (minutes)}}\\\cmidrule(r){2-5}	
			GCD Index   &      36.968$^{***}$&      46.286$^{***}$&     131.935$^{***}$&     158.454$^{***}$\\
            &    (10.589)        &     (7.060)        &    (24.632)        &    (13.397)        \\
Observations&       60348        &       60348        &       60348        &       60348        \\
R-squared   &      575.00        &      569.00        &      575.00        &      569.00        \\
F-Stat      &        0.07        &        0.18        &        0.03        &        0.04        \\
widstat     &                    &                    &      398.31        &     1537.17        \\

			\vspace{0.2cm}
			&\multicolumn{4}{c}{\sc{Panel III. Dependent Variable:  Annual GHG (metric tons)}}\\\cmidrule(r){2-5}
			GCD Index   &       1.317$^{***}$&       0.814$^{***}$&       3.006$^{***}$&       1.432$^{***}$\\
            &     (0.178)        &     (0.100)        &     (0.413)        &     (0.254)        \\
Observations&       60353        &       60353        &       60353        &       60353        \\
R-squared   &      575.00        &      569.00        &      575.00        &      569.00        \\
F-Stat      &        0.11        &        0.52        &       -0.01        &        0.19        \\
widstat     &                    &                    &      398.30        &     1536.44        \\
	
			\\
			\textsl{Controls:}\\
			Distance &  \checkmark & \checkmark & \checkmark & \checkmark\\
			Geography &  & \checkmark && \checkmark  \\
			State \& Metro Fixed Effects &  & \checkmark &  & \checkmark\\
			\\\bottomrule
	\end{tabular}}
	\begin{minipage}{1\linewidth}												
		\scriptsize \textsl{Note.--- Columns 1-2 present OLS estimates of the relationship between GCD (using the continuous measure) and neighborhood outcomes. Column 1 controls for location using distance to the main city within each MSA using 5-km distance bins. Column 2 adds geographic controls, including elevation, slope, soil type (ecological regions), latitude and longitude, and metro area and state fixed effects. Columns 3–4 replicate the specifications from Columns 1–2 but using national design waves as instruments for GCD adoption. The coefficients with *** are significant at the 1\% confidence level; with ** are significant at the 5\% confidence level; and with * are significant at the 10\% confidence level. Robust standard errors, adjusted for clustering by county, are in parentheses. OLS and IV estimates are weighted by neighborhood population.}								
	\end{minipage}	
\end{table}

\begin{table}[!ht]
	\centering
	\caption{\sc{Estimates of Garden Design on Social and Environmental Outcomes using First Principal Component of Four GCD Features}}
    \label{table: pca robust}
	\resizebox{1\textwidth}{!}{\begin{tabular}{L{5cm}C{3cm}C{3cm}C{3cm}C{3cm}C{3cm}}\toprule\toprule
			\vspace{0.5cm}
			&(I) &(II)&(III)&(IV)&(V)\\
			\cmidrule(r){2-3} \cmidrule(r){4-4}  \cmidrule(r){5-6} \\
			&\multicolumn{2}{c}{OLS Estimates} & 	\multicolumn{1}{c}{Propensity Score} & 	\multicolumn{2}{c}{IV Estimates}   \\
			\vspace{0.5cm}
			&\multicolumn{5}{c}{\sc{Panel I. Dependent Variable:  Log Social isolation}}\\\cmidrule(r){2-6}	
			        &                    &                    &                    &                    &                    \\
High-GCD neighborhoods (Top 20\% PCA)&       0.185$^{***}$&       0.172$^{***}$&       0.146$^{***}$&       0.371$^{***}$&       0.409$^{***}$\\
            &     (0.028)        &     (0.027)        &     (0.026)        &     (0.070)        &     (0.066)        \\
Observations&       45226        &       45226        &       45226        &       45226        &       45226        \\
R-squared   &        0.01        &        0.05        &                    &        0.01        &        0.01        \\
F-Stat      &                    &                    &                    &      620.41        &     1283.75        \\

			\vspace{0.2cm}
			&\multicolumn{5}{c}{\sc{Panel II. Dependent Variable:  Daily Time at Home (minutes)}}\\\cmidrule(r){2-6}	
			        &                    &                    &                    &                    &                    \\
High-GCD neighborhoods (Top 20\% PCA)&      12.774$^{***}$&      14.949$^{***}$&      14.641$^{***}$&      57.975$^{***}$&      65.854$^{***}$\\
            &     (3.146)        &     (1.744)        &     (1.963)        &    (10.383)        &     (5.576)        \\
Observations&       60348        &       60348        &       60348        &       60348        &       60348        \\
R-squared   &        0.07        &        0.18        &                    &        0.01        &        0.01        \\
F-Stat      &                    &                    &                    &      741.78        &     1803.85        \\

			\vspace{0.2cm}
			&\multicolumn{5}{c}{\sc{Panel III. Dependent Variable:  Annual GHG (metric tons)}}\\\cmidrule(r){2-6}
			       &                    &                    &                    &                    &                    \\
High-GCD neighborhoods (Top 20\% PCA)&       0.347$^{***}$&       0.186$^{***}$&       0.234$^{***}$&       1.270$^{***}$&       0.526$^{***}$\\
            &     (0.054)        &     (0.031)        &     (0.026)        &     (0.177)        &     (0.109)        \\
Observations&       60353        &       60353        &       60353        &       60353        &       60353        \\
R-squared   &        0.07        &        0.51        &                    &       -0.18        &        0.14        \\
F-Stat      &                    &                    &                    &      741.79        &     1803.49        \\
	
			\\
			\textsl{Controls:}\\
			Distance &  \checkmark & \checkmark & \checkmark & \checkmark & \checkmark\\
			Geography &  & \checkmark & \checkmark & & \checkmark \\
			State \& Metro Fixed Effects &  & \checkmark  & \checkmark & & \checkmark \\
			\\\bottomrule
	\end{tabular}}
	\begin{minipage}{1\linewidth}												
		\scriptsize \textsl{Note.--- This table replicates the baseline analysis reported on table \ref{table: all discrete} using the first principal component of the Garden City Design (GCD) features. The top 20\% of neighborhoods in the PCA distribution are classified as ``high GCD''. Columns 1–2 report OLS estimates controlling for distance to the city center (5-km bins) and, in Column 2, additional covariates including elevation, slope, ecological region, latitude, longitude, and metro area and state fixed effects. Column 3 reports estimates from propensity score matching, using the same set of covariates as Column 2 to estimate the propensity score. Columns 4–5 replicate the same specifications using national design waves as instruments for GCD adoption. The coefficients with *** are significant at the 1\% confidence level; with ** are significant at the 5\% confidence level; and with * are significant at the 10\% confidence level. Robust standard errors, adjusted for clustering by county, are in parentheses. OLS and IV estimates are weighted by neighborhood population.}								
	\end{minipage}	
\end{table}	

\begin{table}[!ht]
	\centering 
	\caption{\sc{Estimates of Garden Design on Social and Environmental Outcomes using Top Tercile of GCD Measure}}
    \label{table: tertile robust}

	\resizebox{1\textwidth}{!}{\begin{tabular}{L{6.5cm}C{2cm}C{2cm}C{2cm}C{2cm}C{2cm}}\toprule\toprule
			\vspace{0.5cm}
			&(I) &(II)&(III)&(IV)&(V)\\
			\cmidrule(r){2-3} \cmidrule(r){4-4}  \cmidrule(r){5-6} \\
			&\multicolumn{2}{c}{OLS Estimates} & 	\multicolumn{1}{c}{Propensity Score} & 	\multicolumn{2}{c}{IV Estimates}   \\
			\vspace{0.5cm}
			&\multicolumn{5}{c}{\sc{Panel I. Dependent Variable:  Log Social isolation}}\\\cmidrule(r){2-6}	
			      &                    &                    &                    &                    &                    \\
High-GCD neighborhoods (Top 33\%)&       0.146$^{***}$&       0.128$^{***}$&       0.113$^{***}$&       0.303$^{***}$&       0.343$^{***}$\\
            &     (0.026)        &     (0.021)        &     (0.019)        &     (0.061)        &     (0.058)        \\
Observations&       45226        &       45226        &       45226        &       45226        &       45226        \\
R-squared   &        0.01        &        0.05        &                    &        0.01        &        0.01        \\
F-Stat      &                    &                    &                    &      782.38        &     2125.30        \\

			\vspace{0.2cm}
			&\multicolumn{5}{c}{\sc{Panel II. Dependent Variable:  Daily Time at Home (minutes)}}\\\cmidrule(r){2-6}	
			       &                    &                    &                    &                    &                    \\
High-GCD neighborhoods (Top 33\%)&      12.467$^{***}$&      13.497$^{***}$&      12.974$^{***}$&      50.621$^{***}$&      58.496$^{***}$\\
            &     (2.994)        &     (1.896)        &     (1.973)        &     (9.290)        &     (5.216)        \\
Observations&       60348        &       60348        &       60348        &       60348        &       60348        \\
R-squared   &        0.07        &        0.18        &                    &        0.02        &        0.01        \\
F-Stat      &                    &                    &                    &      798.62        &     2492.29        \\

			\vspace{0.2cm}
			&\multicolumn{5}{c}{\sc{Panel III. Dependent Variable:  Annual GHG (metric tons)}}\\\cmidrule(r){2-6}
			        &                    &                    &                    &                    &                    \\
High-GCD neighborhoods (Top 33\%)&       0.360$^{***}$&       0.205$^{***}$&       0.236$^{***}$&       1.151$^{***}$&       0.516$^{***}$\\
            &     (0.052)        &     (0.028)        &     (0.025)        &     (0.153)        &     (0.096)        \\
Observations&       60353        &       60353        &       60353        &       60353        &       60353        \\
R-squared   &        0.08        &        0.51        &                    &       -0.15        &        0.14        \\
F-Stat      &                    &                    &                    &      798.53        &     2490.97        \\
	
			\\
			\textsl{Controls:}\\
			Distance &  \checkmark & \checkmark & \checkmark & \checkmark & \checkmark\\
			Geography &  & \checkmark & \checkmark & & \checkmark \\
			State \& Metro Fixed Effects &  & \checkmark  & \checkmark & & \checkmark \\
			\\\bottomrule 
	\end{tabular}}
	\begin{minipage}{1\linewidth}												
		\scriptsize \textsl{Note.--- This table replicates the baseline analysis reported on table \ref{table: all discrete}, with the top 33\% of neighborhoods in the GCD distribution classified as ``high GCD''. Columns 1–2 report OLS estimates controlling for distance to the city center (5-km bins) and, in Column 2, additional covariates including elevation, slope, ecological region, latitude, longitude, and metro area and state fixed effects. Column 3 reports estimates from propensity score matching, using the same set of covariates as Column 2 to estimate the propensity score. Columns 4–5 replicate the same specifications using national design waves as instruments for GCD adoption. The coefficients with *** are significant at the 1\% confidence level; with ** are significant at the 5\% confidence level; and with * are significant at the 10\% confidence level. Robust standard errors, adjusted for clustering by county, are in parentheses. OLS and IV estimates are weighted by neighborhood population.}								
	\end{minipage}	
\end{table}		

\begin{table}[!ht]
	\centering
	\caption{\sc{OLS estimates controlling for demographics, historical covariates, and work-from-home.}}
	\label{table: ols discrete robustness}
	\resizebox{1\textwidth}{!}{\begin{tabular}{L{5cm}C{2.5cm}C{2.5cm}C{2.5cm}C{2.5cm}C{2.5cm}C{2.5cm}}\toprule\toprule
			\vspace{0.2cm}
			&(I) &(II)&(III)&(IV) &(V) &(VI) \\
			&\multicolumn{5}{c}{\sc{Panel I. Dependent Variable: Log Social isolation}}\\\cmidrule(r){2-7}
			High-GCD neighborhoods (Top 20\%)&       0.207$^{***}$&       0.192$^{***}$&       0.185$^{***}$&       0.093$^{***}$&       0.277$^{***}$&       0.271$^{***}$\\
            &     (0.026)        &     (0.026)        &     (0.027)        &     (0.028)        &     (0.026)        &     (0.026)        \\
Observations&       45226        &       45209        &       45209        &       45164        &       45209        &       45199        \\
R-squared   &        0.01        &        0.05        &        0.05        &        0.08        &        0.07        &        0.07        \\

			\vspace{0.5cm}
			&\multicolumn{6}{c}{\sc{Panel III. Dependent Variable: Daily Time at Home (minutes)}}\\\cmidrule(r){2-7}
			High-GCD neighborhoods (Top 20\%)&      12.920$^{***}$&      15.052$^{***}$&      14.272$^{***}$&      12.081$^{***}$&      14.910$^{***}$&      14.417$^{***}$\\
            &     (3.086)        &     (1.671)        &     (1.689)        &     (1.837)        &     (1.551)        &     (1.642)        \\
Observations&       60348        &       60335        &       60335        &       60199        &       60335        &       51676        \\
R-squared   &        0.07        &        0.18        &        0.18        &        0.21        &        0.18        &        0.18        \\

			\vspace{0.5cm}
			&\multicolumn{6}{c}{\sc{Panel II. Dependent Variable: Annual GHG (metric tons)}}\\\cmidrule(r){2-7}
			High-GCD neighborhoods (Top 20\%)&       0.344$^{***}$&       0.185$^{***}$&       0.175$^{***}$&       0.143$^{***}$&       0.084$^{***}$&       0.097$^{***}$\\
            &     (0.054)        &     (0.033)        &     (0.032)        &     (0.032)        &     (0.028)        &     (0.027)        \\
Observations&       60353        &       60340        &       60340        &       60203        &       60340        &       51679        \\
R-squared   &        0.07        &        0.50        &        0.52        &        0.54        &        0.57        &        0.57        \\

			\\
			\textsl{Controls:}\\
			Distance &  \checkmark & \checkmark & \checkmark & \checkmark & \checkmark   & \checkmark\\
			Geography &  & \checkmark & \checkmark & \checkmark & \checkmark  & \checkmark\\
			State \& Metro Fixed Effects &  & \checkmark & \checkmark & \checkmark & \checkmark  & \checkmark\\
			Historical Building Density (1900)  &  & & \checkmark & \checkmark & \checkmark  & \checkmark\\
			Demographics (2000) &  & &  & \checkmark & \checkmark   & \checkmark\\
			Population Density (2000) &  & &  &  & \checkmark  & \checkmark \\
			Work From Home (Tract level) &  & &  &  & & \checkmark  \\			
			\\\bottomrule
	\end{tabular}}
	\begin{minipage}{1\linewidth}												
		\scriptsize \textsl{Note.---The table presents OLS estimates of the relationship between GCD (using the discrete version) and neighborhood outcomes.  Column 1 controls for location using distance to the main city within each MSA. Column 2 adds geographic controls, including elevation, slope, soil type (ecological regions), and latitude and longitude, plus metro area and state fixed-effects. Column 3 further controls for historical building density in 1900 using the data from \citet{leyk_2018}. Column 4 controls for the share of white population, share of married households, share of people with some college, median age of the population, and the log average per capita income. Column 5 controls for  population density in 2000.  Note that the demographic composition and density of a neighborhood are potentially a function of its design. When this is the case, these covariates are ``bad controls'' \citep{angrist_2009, angrist_2015}, and their inclusion can lead to other forms of bias on my estimates of $\beta$. However, the fact that the estimates in column 3 are similar to those in columns 4 and 5 suggests that any such bias is small and that the true effect of GCD is between these estimates range. Column 6 controls for the share of workers in the broader Census tract who reported working from home. The coefficients with *** are significant at the 1\% confidence level; with ** are significant at the 5\% confidence level; and with * are significant at the 10\% confidence level. Robust standard errors, adjusted for clustering by county, are in parentheses. All estimates are weighted by neighborhood population.}								
	\end{minipage}	
\end{table}	

\begin{table}[!ht]
	\centering
	\caption{\sc{Complementary Outcomes}}
     \label{table: complementary outcomes}
	\resizebox{1\textwidth}{!}{\begin{tabular}{L{7cm}C{3cm}C{3cm}C{3cm}C{3cm}C{3cm}}\toprule\toprule
			\vspace{0.5cm}
			&(I) &(II)&(III)&(IV) &(V)\\
			\cmidrule(r){2-3} \cmidrule(r){4-4}  \cmidrule(r){5-6} \\
			&\multicolumn{2}{c}{OLS Estimates} & 	\multicolumn{1}{c}{Propensity Score} & 	\multicolumn{2}{c}{IV Estimates}   \\
			\vspace{0.5cm}	
			&\multicolumn{5}{c}{\sc{Panel I. Dependent Variable: Annual GHG Emissions from Commuting (metric tons)
			}}\\\cmidrule(r){2-6}	
			        &                    &                    &                    &                    &                    \\
High-GCD neighborhoods (Top 20\%)&       0.244$^{***}$&       0.138$^{***}$&       0.177$^{***}$&       0.910$^{***}$&       0.376$^{***}$\\
            &     (0.045)        &     (0.030)        &     (0.025)        &     (0.156)        &     (0.101)        \\
Observations&       60353        &       60353        &       60353        &       60353        &       60353        \\
R-squared   &        0.06        &        0.46        &                    &       -0.12        &        0.14        \\
F-Stat      &                    &                    &                    &      671.98        &     1697.57        \\

            \vspace{0.2cm}	
            &\multicolumn{5}{c}{\sc{Panel II. Dependent Variable: Annual GHG Emissions from Non-Commuting (km)
			}}\\\cmidrule(r){2-6}	
			    &                    &                    &                    &                    &                    \\
High-GCD neighborhoods (Top 20\%)&       0.100$^{***}$&       0.047$^{***}$&       0.060$^{***}$&       0.369$^{***}$&       0.143$^{***}$\\
            &     (0.015)        &     (0.004)        &     (0.004)        &     (0.044)        &     (0.012)        \\
Observations&       60353        &       60353        &       60353        &       60353        &       60353        \\
R-squared   &        0.08        &        0.64        &                    &       -0.37        &        0.10        \\
F-Stat      &                    &                    &                    &      671.98        &     1697.57        \\

            \vspace{0.2cm}	
			&\multicolumn{5}{c}{\sc{Panel III. Dependent Variable: \% of population owning 2+
					automobiles
			}}\\\cmidrule(r){2-6}	
			       &                    &                    &                    &                    &                    \\
High-GCD neighborhoods (Top 20\%)&       0.088$^{***}$&       0.067$^{***}$&       0.078$^{***}$&       0.285$^{***}$&       0.173$^{***}$\\
            &     (0.012)        &     (0.005)        &     (0.005)        &     (0.043)        &     (0.021)        \\
Observations&       60353        &       60353        &       60353        &       60353        &       60353        \\
R-squared   &        0.14        &        0.31        &                    &       -0.01        &        0.14        \\
F-Stat      &                    &                    &                    &      671.98        &     1697.57        \\

			\vspace{0.2cm}	
			&\multicolumn{5}{c}{\sc{Panel IV. Dependent Variable: Point of Interest Density
			}}\\\cmidrule(r){2-6}	
			        &                    &                    &                    &                    &                    \\
High-GCD neighborhoods (Top 20\%)&      -0.581$^{***}$&      -0.424$^{***}$&      -0.524$^{***}$&      -1.526$^{***}$&      -1.200$^{***}$\\
            &     (0.066)        &     (0.037)        &     (0.036)        &     (0.170)        &     (0.114)        \\
Observations&       57978        &       57978        &       58042        &       57978        &       57978        \\
R-squared   &        0.06        &        0.17        &                    &       -0.05        &        0.03        \\
F-Stat      &                    &                    &                    &      671.50        &     1727.40        \\

			\vspace{0.2cm}	
			&\multicolumn{5}{c}{\sc{Panel V. Dependent Variable: Walkscore
			}}\\\cmidrule(r){2-6}	
			        &                    &                    &                    &                    &                    \\
High-GCD neighborhoods (Top 20\%)&      -1.277$^{***}$&      -1.081$^{***}$&      -1.216$^{***}$&      -4.001$^{***}$&      -3.600$^{***}$\\
            &     (0.119)        &     (0.071)        &     (0.086)        &     (0.339)        &     (0.225)        \\
Observations&       60353        &       60353        &       60353        &       60353        &       60353        \\
R-squared   &        0.07        &        0.28        &                    &       -0.09        &        0.00        \\
F-Stat      &                    &                    &                    &      671.98        &     1697.57        \\
 \\           
			\textsl{Controls:}\\
			Distance &  \checkmark & \checkmark & \checkmark & \checkmark & \checkmark\\
			Geography &  & \checkmark & \checkmark & & \checkmark \\
			State \& Metro Fixed Effects &  & \checkmark  & \checkmark & & \checkmark \\
			\\\bottomrule
	\end{tabular}}
	\begin{minipage}{1\linewidth}												
		\scriptsize \textsl{Note.---This table examines complementary outcomes associated with travel behavior. Columns 1–2 present OLS estimates of the relationship between GCD (using the discrete measure) and each outcome. Column 1 controls for location using distance to the main city within each MSA (5-km bins), and Column 2 adds geographic covariates including elevation, slope, ecological region, latitude, longitude, and metropolitan area and state fixed effects. Column 3 reports estimates from propensity score matching, using the same set of covariates as Column 2 to estimate the propensity score. Columns 4–5 replicate these specifications using national design waves as instruments for GCD adoption.  Robust standard errors, clustered by county, are shown in parentheses. The coefficients with *** are significant at the 1\% confidence level; with ** at 5\%; and with * at 10\%. OLS and IV estimates are weighted by neighborhood population.}								
	\end{minipage}	
\end{table}	

\begin{table}[!ht]
	\centering
	\caption{\sc{Balance of Covariates}}
	\label{table: covariate balance}
	\resizebox{1\textwidth}{!}{\begin{tabular}{L{7cm}C{3cm}C{3cm}C{3cm}C{3cm}C{3cm}}\toprule\toprule
			&\multicolumn{5}{c}{\sc{Samples}}\\\cmidrule(r){2-6}
			\vspace{0.2cm}
 & \begin{tabular}{c} High-GCD \\ Unweighted \end{tabular} 
& \begin{tabular}{c} Low-GCD \\ Unweighted \end{tabular} 
& \begin{tabular}{c} High-GCD \\ Weighted \end{tabular} 
& \begin{tabular}{c} Low-GCD \\ Weighted \end{tabular} 
& \begin{tabular}{c} T-test \end{tabular} \\
			\\
			&\multicolumn{5}{c}{\sc{Panel I. Controls}}\\\cmidrule(r){2-6}
			\multirow{2}{7cm}{Distance to City Center (km)\dotfill}&   26.683 &    20.607 &    22.143 &    21.740 &     0.403 \\  
			& (   21.031) &  (   21.493) &  (    0.268) &  (    0.098) &  [$ p=$    0.712] \\ 
			\multirow{2}{7cm}{Elevation\dotfill}&  216.548 &   195.451 &   199.717 &   198.750 &     0.966 \\  
			& (  367.456) &  (  318.610) &  (    3.228) &  (    1.583) &  [$ p=$    0.942] \\ 
			\multirow{2}{7cm}{Slope\dotfill}&    0.602 &     0.484 &     0.523 &     0.511 &     0.012 \\  
			& (    0.691) &  (    0.549) &  (    0.006) &  (    0.003) &  [$ p=$    0.546] \\ 
			\multirow{2}{7cm}{Latitude\dotfill}&   36.854 &    38.189 &    37.896 &    37.915 &    -0.019 \\  
			& (    5.116) &  (    4.794) &  (    0.061) &  (    0.023) &  [$ p=$    0.954] \\ 
			\multirow{2}{7cm}{Longitude\dotfill}&  -95.764 &   -93.413 &   -94.117 &   -93.887 &    -0.230 \\  
			& (   18.585) &  (   17.752) &  (    0.204) &  (    0.084) &  [$ p=$    0.885] \\ 
			\multirow{2}{7cm}{Share in Midwest\dotfill}&    0.119 &     0.230 &     0.206 &     0.207 &    -0.000 \\  
			& (    0.324) &  (    0.421) &  (    0.007) &  (    0.002) &  [$ p=$    0.990] \\ 
			\multirow{2}{7cm}{Share in Northeast\dotfill}&    0.189 &     0.249 &     0.238 &     0.238 &    -0.001 \\  
			& (    0.391) &  (    0.433) &  (    0.006) &  (    0.002) &  [$ p=$    0.973] \\ 
			\multirow{2}{7cm}{Share in South\dotfill}&    0.302 &     0.212 &     0.225 &     0.229 &    -0.004 \\  
			& (    0.459) &  (    0.408) &  (    0.004) &  (    0.002) &  [$ p=$    0.854] \\ 			
			\multirow{2}{7cm}{Share in West\dotfill}&    0.390 &     0.309 &     0.331 &     0.326 &     0.005 \\  
			& (    0.488) &  (    0.462) &  (    0.005) &  (    0.002) &  [$ p=$    0.898] \\ 
			\\\bottomrule
	\end{tabular}}
	\begin{minipage}{1\linewidth}											
	\scriptsize \textsl{Note.---Columns 1 and 2 present unweighted means for high-GCD neighborhoods (top 20\% of the GCD distribution) and low-GCD neighborhoods (bottom 80\%), respectively. Columns 3 and 4 report the corresponding weighted means after applying inverse propensity weighting. }								
\end{minipage}	
\end{table}

\begin{table}[!ht]
	\centering
	\caption{\sc{Variance decomposition of the prevalence of GCD across US neighborhoods.}}
	\label{table: variance decomposition}
	\resizebox{1\textwidth}{!}{\begin{tabular}{L{9cm}C{2cm}C{2cm}
    }\toprule\toprule
			\vspace{0.2cm}
			&(I) &(II)\\
			\vspace{0.2cm}
			&\multicolumn{2}{c}{\sc{Panel I. Dependent Variable: GCD Index (discrete)}}\\\cmidrule(r){2-3}
			Observations&       58673        &       58673      \\
Total R-squared&        0.20        &        0.26       \\
GCD National Waves&        0.19        &        0.21        \\
All local determinants&        0.01        &        0.05        \\	
			&\multicolumn{2}{c}{\sc{Panel II. Dependent Variable: GCD Index (continuous) }}\\\cmidrule(r){2-3}
			Observations&       58673        &       58673       \\
Total R-squared&        0.31        &        0.42       \\
GCD National Waves&        0.28        &        0.29        \\
All local determinants&        0.02        &        0.12        \\ \\
            \textsl{Local covariates included:}\\
			Distance &  \checkmark & \checkmark \\
			Geography &  & \checkmark \\
			State \& Metro Fixed Effects &  & \checkmark \\
			\\\bottomrule
	\end{tabular}}
	\begin{minipage}{1\linewidth}												
		\scriptsize \textsl{Note.---} The Table decomposes the variance in GCD (using the discrete measure in Panel I and the continuous measure in Panel II). Column 1 explains GCD using 5-year historical cohorts and distance to the main city in the neighborhood's MSA. Column 2 adds local geographic characteristics (elevation, slope, soil type (ecological regions), and latitude and longitude), MSA and State fixed effects. The decomposition is based on the regression model
        \begin{align*}
    \textrm{Garden City Design}_{ims} = \textrm{Cohort}_{ims} + G(d(i)) + \eta_{m} + \nu_s + \rho\; X_{ims} + \upsilon_{ims}. 
\end{align*}
        
	\end{minipage}	
\end{table}

\begin{table}[!ht]
	\centering
	\caption{\sc{OLS and IV estimates controlling for the role of residential zoning}}
	\label{table: zoning}
	\resizebox{1\textwidth}{!}{\begin{tabular}{L{7.5cm}C{3.2cm}C{3.2cm}C{3.2cm}C{3.2cm}}\toprule\toprule
			\vspace{0.5cm}
			&(I) &(II)&(III)&(IV) \\
			\cmidrule(r){2-3} \cmidrule(r){4-5} \\
			&\multicolumn{2}{c}{OLS Estimates} & 	\multicolumn{2}{c}{IV Estimates}  \\
			\vspace{0.5cm}
			&\multicolumn{4}{c}{\sc{Panel I. Dependent Variable:  Log Social isolation}}\\\cmidrule(r){2-5}	
			High-GCD neighborhoods (Top 20\%)&       0.192$^{***}$&       0.200$^{***}$&       0.403$^{***}$&       0.672$^{***}$\\
            &     (0.026)        &     (0.033)        &     (0.066)        &     (0.099)        \\
Share of Plots Zoned Residential&                    &       1.016$^{***}$&                    &       1.051$^{***}$\\
            &                    &     (0.096)        &                    &     (0.092)        \\
Observations&       45226        &       27335        &       45226        &       27335        \\
R-squared   &        0.05        &        0.10        &        0.02        &        0.05        \\
F-Stat      &                    &                    &     1242.11        &      752.17        \\

			\vspace{0.2cm}
			&\multicolumn{4}{c}{\sc{Panel II. Dependent Variable:  Daily Time at Home (minutes)}}\\\cmidrule(r){2-5}	
			High-GCD neighborhoods (Top 20\%)&      15.052$^{***}$&      17.119$^{***}$&      64.367$^{***}$&      77.185$^{***}$\\
            &     (1.671)        &     (2.093)        &     (5.553)        &     (6.390)        \\
Share of Plots Zoned Residential&                    &      23.550$^{***}$&                    &      29.024$^{***}$\\
            &                    &     (5.572)        &                    &     (5.014)        \\
Observations&       60348        &       36921        &       60348        &       36921        \\
R-squared   &        0.18        &        0.21        &        0.01        &       -0.03        \\
F-Stat      &                    &                    &     1697.81        &     1016.23        \\

			\vspace{0.2cm}
			&\multicolumn{4}{c}{\sc{Panel III. Dependent Variable:  Annual GHG (metric tons)}}\\\cmidrule(r){2-5}
			High-GCD neighborhoods (Top 20\%)&       0.185$^{***}$&       0.141$^{***}$&       0.519$^{***}$&       0.466$^{***}$\\
            &     (0.033)        &     (0.040)        &     (0.108)        &     (0.144)        \\
Share of Plots Zoned Residential&                    &       0.071        &                    &       0.100        \\
            &                    &     (0.087)        &                    &     (0.093)        \\
Observations&       60353        &       36923        &       60353        &       36923        \\
R-squared   &        0.50        &        0.45        &        0.14        &        0.04        \\
F-Stat      &                    &                    &     1697.57        &     1016.00        \\
	
				\\
			\textsl{Controls:}\\
			Distance &  \checkmark & \checkmark & \checkmark & \checkmark \\
			Geography &  \checkmark & \checkmark & \checkmark & \checkmark \\
			Metro \& State Fixed Effects  &  \checkmark & \checkmark & \checkmark & \checkmark \\
			Zoning  & & \checkmark && \checkmark \\
			\\\bottomrule
	\end{tabular}}
	\begin{minipage}{1\linewidth}												
		\scriptsize \textsl{Note.---
        The table explores the robustness of the relationship between GCD and the main outcomes when controlling for the prevalence of residential zoning.  
        Columns 1-2 present OLS estimates. Column 1 is the baseline specification, controlling for distance to city center, geography and metro and state fixed effects. Column 2 adds the share of residential plots in the neighborhood as a control. Columns 3-4 replicate these specifications using national design waves as instruments for GCD adoption.  Robust standard errors, clustered by county, are shown in parentheses. The coefficients with *** are significant at the 1\% confidence level; with ** at 5\%; and with * at 10\%. OLS and IV estimates are weighted by neighborhood population.}								
	\end{minipage}	
\end{table}

\begin{table}[!ht]
	\centering
	\caption{\sc{OLS and IV estimates controlling for age effects}}
	\label{table: age effects}
	\resizebox{1\textwidth}{!}{\begin{tabular}{L{5cm}C{2.8cm}C{2.8cm}C{2.8cm}C{2.8cm}C{2.8cm}C{2.8cm}}\toprule\toprule
			\vspace{0.5cm}
			&(I) &(II)&(III)&(IV)&(V)&(VI) \\
			\cmidrule(r){2-3} \cmidrule(r){4-5}  \cmidrule(r){6-7} \\
			&\multicolumn{2}{c}{Age Effects (1800--1875 sample) } & 	\multicolumn{2}{c}{IV Estimates (full sample)} & 	\multicolumn{2}{c}{OLS Estimates (full sample)}  \\
			\vspace{0.5cm}
			&\multicolumn{6}{c}{\sc{Panel I. Dependent Variable:  Log Social isolation}}\\\cmidrule(r){2-7}	
			Age Effect  &      -0.003$^{***}$&      -0.003$^{***}$&                    &                    &                    &                    \\
            &     (0.001)        &     (0.001)        &                    &                    &                    &                    \\
High-GCD neighborhoods (Top 20\%)&                    &                    &       0.151$^{**}$ &       0.156$^{**}$ &       0.159$^{***}$&       0.143$^{***}$\\
            &                    &                    &     (0.066)        &     (0.069)        &     (0.027)        &     (0.027)        \\
Observations&        2720        &        2720        &       43977        &       43977        &       43977        &       43977        \\
R-squared   &        0.11        &        0.13        &        0.01        &        0.02        &        0.04        &        0.05        \\
F-Stat      &                    &                    &     1104.90        &     1122.98        &                    &                    \\

			\vspace{0.2cm}
			&\multicolumn{6}{c}{\sc{Panel II. Dependent Variable:  Daily Time at Home (minutes)}}\\\cmidrule(r){2-7}	
			Age Effect  &      -0.300$^{***}$&      -0.319$^{***}$&                    &                    &                    &                    \\
            &     (0.081)        &     (0.077)        &                    &                    &                    &                    \\
High-GCD neighborhoods (Top 20\%)&                    &                    &      44.811$^{***}$&      37.104$^{***}$&      10.969$^{***}$&       9.192$^{***}$\\
            &                    &                    &     (5.809)        &     (5.301)        &     (1.760)        &     (1.614)        \\
Observations&        3509        &        3509        &       58624        &       58624        &       58624        &       58624        \\
R-squared   &        0.15        &        0.17        &        0.01        &        0.03        &        0.15        &        0.15        \\
F-Stat      &                    &                    &     1566.27        &     1611.95        &                    &                    \\

			\vspace{0.2cm}	
			&\multicolumn{6}{c}{\sc{Panel III. Dependent Variable:  Annual GHG (metric tons)}}\\\cmidrule(r){2-7}
	    	Age Effect  &      -0.002$^{***}$&      -0.002$^{***}$&                    &                    &                    &                    \\
            &     (0.001)        &     (0.001)        &                    &                    &                    &                    \\
High-GCD neighborhoods (Top 20\%)&                    &                    &       0.224$^{*}$  &       0.293$^{***}$&       0.147$^{***}$&       0.145$^{***}$\\
            &                    &                    &     (0.118)        &     (0.110)        &     (0.034)        &     (0.033)        \\
Observations&        3510        &        3510        &       58628        &       58628        &       58628        &       58628        \\
R-squared   &        0.46        &        0.56        &        0.08        &        0.17        &        0.44        &        0.50        \\
F-Stat      &                    &                    &     1566.16        &     1611.84        &                    &                    \\
	
			\\
			\textsl{Controls:}\\
            Distance &  \checkmark & \checkmark & \checkmark & \checkmark & \checkmark & \checkmark \\
			Metro \& State Fixed Effects & & \checkmark && \checkmark &&  \checkmark \\
            Geography &  &\checkmark && \checkmark && \checkmark  \\
			\\\bottomrule
	\end{tabular}}
	\begin{minipage}{1\linewidth}												
		\scriptsize \textsl{Note.--- Columns 1 and 2 report estimates of age effects, from the sample of neighborhoods developed before 1875. Column 1 controls for distance to city center, using 5-km dummies. Column 2 adds geographic controls, including elevation, slope, soil type (ecological regions), latitude and longitude, and metro area and state fixed effects.
        The remaining columns present IV and OLS estimates that age-adjust the outcomes (computing them as $\textrm{Outcome}_{ims} - \widehat\beta_a\;\textrm{Age}_{ims},$ where $\widehat\beta_a$ are the estimates from Columns 1 and 2, respectively) using the sample of all neighborhoods and the same two specifications. The coefficients with *** are significant at the 1\% confidence level; with ** are significant at the 5\% confidence level; and with * are significant at the 10\% confidence level. Robust standard errors, adjusted for clustering by county, are in parentheses. OLS and IV estimates are weighted by neighborhood population.}								
	\end{minipage}	
\end{table}

\begin{table}[!ht]
	\centering
	\caption{\sc{Robustness checks for IV estimates of the relationship between GCD and neighborhood outcomes}}
	\label{table: iv robustness}
	\resizebox{1\textwidth}{!}{\begin{tabular}{L{5cm}C{3cm}C{3cm}C{3cm}C{3cm}}\toprule\toprule
			\vspace{0.5cm}
			&(I) &(II)&(III)&(IV) \\
			\cmidrule(r){2-3} \cmidrule(r){4-5} \\
            &\multicolumn{2}{c}{Leave State Out} &
            \multicolumn{2}{c}{%
              \begin{tabular}{@{}c@{}}
                Controlling for Development Periods\\
                (1810-1875; 1875--1950; 1950-2020)
              \end{tabular}
            } \\
			\vspace{0.5cm}
			&\multicolumn{4}{c}{\sc{Panel I. Dependent Variable: Log Social isolation}}\\\cmidrule(r){2-5}
			High-GCD neighborhoods (Top 20\%)&       0.375$^{***}$&       0.403$^{***}$&       0.490$^{***}$&       0.538$^{***}$\\
            &     (0.072)        &     (0.066)        &     (0.096)        &     (0.083)        \\
Observations&       45226        &       45226        &       45226        &       45226        \\
R-squared   &        0.01        &        0.02        &        0.01        &        0.01        \\
F-Stat      &      551.73        &     1242.11        &      388.18        &      926.24        \\

			&\multicolumn{4}{c}{\sc{Panel II. Dependent Variable: Daily Time at Home (minutes)}}\\\cmidrule(r){2-5}
			High-GCD neighborhoods (Top 20\%)&      58.133$^{***}$&      64.367$^{***}$&      48.098$^{***}$&      60.994$^{***}$\\
            &    (10.504)        &     (5.553)        &     (9.086)        &     (6.021)        \\
Observations&       60348        &       60348        &       60348        &       60348        \\
R-squared   &        0.01        &        0.01        &        0.04        &        0.02        \\
F-Stat      &      671.96        &     1697.81        &      506.37        &     1163.49        \\
	
			&\multicolumn{4}{c}{\sc{Panel III. Dependent Variable: Annual GHG (metric tons)}}\\\cmidrule(r){2-5}
			High-GCD neighborhoods (Top 20\%)&       1.279$^{***}$&       0.519$^{***}$&       0.734$^{***}$&       0.276$^{**}$ \\
            &     (0.178)        &     (0.108)        &     (0.181)        &     (0.110)        \\
Observations&       60353        &       60353        &       60353        &       60353        \\
R-squared   &       -0.19        &        0.14        &        0.06        &        0.19        \\
F-Stat      &      671.98        &     1697.57        &      506.53        &     1163.47        \\

			\\
			\textsl{Controls:}\\
			Distance &  \checkmark &  \checkmark &  \checkmark &  \checkmark\\
			Geography &  & \checkmark &  & \checkmark\\
			State \& Metro Fixed Effects &  & \checkmark   &  & \checkmark \\
			\\\bottomrule
	\end{tabular}}
	\begin{minipage}{1\linewidth}												
		\scriptsize \textsl{Note.--- The table presents IV estimates of the relationship between GCD and neighborhood outcomes. Columns 1 and 2 reproduce the specifications from Columns 3 and 4 in Table \ref{table: all discrete} but construct the instrument as the average GCD prevalence outside the state of a neighborhood. This check exploits waves in the rest of the country as an instrument for the use of GCD among neighborhoods in a state. Columns 3 and 4 reproduce the specifications from Columns 3 and 4 in Table \ref{table: all discrete} but control for dummies capturing the three phases of GCD use in the US (1810-1875, 1875-1950, 1950-2020). These dummies account for other changes in the development of neighborhoods that varied slowly across these periods.  The coefficients with *** are significant at the 1\% confidence level; with ** are significant at the 5\% confidence level; and with * are significant at the 10\% confidence level. Robust standard errors, adjusted for clustering by county, are in parentheses. All estimates are weighted by neighborhood population.}		\end{minipage}	
\end{table}	

\begin{table}[!ht]
	\centering
	\caption{\sc{OLS and IV estimates of the relationship between GCD and neighborhood outcomes restricted to low-migration neighborhoods}}
	\label{table: low migration}
	\resizebox{1\textwidth}{!}{\begin{tabular}{L{6.5cm}C{3.5cm}C{3.5cm}C{3.5cm}C{3.5cm}}\toprule\toprule
			\vspace{0.5cm}
			&(I) &(II)&(III)&(IV) \\
			\cmidrule(r){2-3} \cmidrule(r){4-5} \\
			&\multicolumn{2}{c}{OLS Estimates} & 	\multicolumn{2}{c}{IV Estimates}  \\
			\vspace{0.5cm}
			&\multicolumn{4}{c}{\sc{Panel I. Dependent Variable: Log Social isolation}}\\\cmidrule(r){2-5}
			High-GCD neighborhoods (Top 20\%)&       0.222$^{***}$&       0.193$^{***}$&       0.435$^{***}$&       0.380$^{***}$\\
            &     (0.020)        &     (0.020)        &     (0.052)        &     (0.053)        \\
Observations&       22659        &       22659        &       22659        &       22659        \\
R-squared   &        0.04        &        0.05        &        0.01        &        0.02        \\
F-Stat      &                    &                    &     3820.35        &     3661.36        \\

			&\multicolumn{4}{c}{\sc{Panel II. Dependent Variable: Daily Time at Home (minutes)}}\\\cmidrule(r){2-5}
			High-GCD neighborhoods (Top 20\%)&      17.070$^{***}$&      14.650$^{***}$&      72.565$^{***}$&      60.296$^{***}$\\
            &     (1.075)        &     (1.076)        &     (2.835)        &     (2.842)        \\
Observations&       30991        &       30991        &       30991        &       30991        \\
R-squared   &        0.15        &        0.17        &       -0.01        &        0.04        \\
F-Stat      &                    &                    &     5730.62        &     5533.24        \\
	
			&\multicolumn{4}{c}{\sc{Panel III. Dependent Variable: Annual GHG (metric tons)}}\\\cmidrule(r){2-5}
			High-GCD neighborhoods (Top 20\%)&       0.183$^{***}$&       0.179$^{***}$&       0.501$^{***}$&       0.506$^{***}$\\
            &     (0.009)        &     (0.009)        &     (0.023)        &     (0.023)        \\
Observations&       30992        &       30992        &       30992        &       30992        \\
R-squared   &        0.50        &        0.54        &        0.05        &        0.13        \\
F-Stat      &                    &                    &     5730.80        &     5533.42        \\

			\\
			\textsl{Controls:}\\
			Distance &  \checkmark &  \checkmark &  \checkmark &  \checkmark\\
			Geography &  & \checkmark &  & \checkmark\\
			State \& Metro Fixed Effects &  & \checkmark   &  & \checkmark \\
			\\\bottomrule
	\end{tabular}}
	\begin{minipage}{1\linewidth}												
		\scriptsize \textsl{Note.---The estimates in this table are for the subset of neighborhoods located in low-migration counties (defined as the bottom 50\% with lower migration rates, according to IRS data). Columns 1 and 3 control for 5-km distance bins to the main city in the neighborhood's MSA. Columns 2 and 4 include additional controls for geography, including elevation, slope, soil type (ecological regions), latitude, longitude, and include metro area and state fixed effects. The coefficients with *** are significant at the 1\% confidence level; with ** are significant at the 5\% confidence level; and with * are significant at the 10\% confidence level. Robust standard
errors, adjusted for clustering by county, are in parentheses. OLS and IV estimates are weighted by neighborhood population.
		}								
	\end{minipage}	
\end{table}

\begin{table}[!ht]
	\centering
	\caption{\sc{Estimates of Garden Design on Social and Environmental Outcomes at the Census Tract Level}\label{table: tract level}}
	\resizebox{1\textwidth}{!}{\begin{tabular}{L{6.5cm}C{2cm}C{2cm}C{2cm}C{2cm}C{2cm}}\toprule\toprule
			\vspace{0.5cm}
			&(I) &(II)&(III)&(IV)&(V)\\
			\cmidrule(r){2-3} \cmidrule(r){4-4}  \cmidrule(r){5-6} \\
			&\multicolumn{2}{c}{OLS Estimates} & 	\multicolumn{1}{c}{Propensity Score} & 	\multicolumn{2}{c}{IV Estimates}   \\
			\vspace{0.5cm}
			&\multicolumn{5}{c}{\sc{Panel I. Dependent Variable:  Log Social isolation}}\\\cmidrule(r){2-6}	
			       &                    &                    &                    &                    &                    \\
High-GCD neighborhoods (Top 20\%)&       0.187$^{***}$&       0.183$^{***}$&       0.216$^{***}$&       0.317$^{***}$&       0.395$^{***}$\\
            &     (0.025)        &     (0.024)        &     (0.028)        &     (0.056)        &     (0.051)        \\
Observations&       14978        &       14978        &       14978        &       14978        &       14978        \\
R-squared   &        0.03        &        0.11        &                    &        0.03        &        0.04        \\
F-Stat      &                    &                    &                    &      365.67        &      943.19        \\

			\vspace{0.2cm}
			&\multicolumn{5}{c}{\sc{Panel II. Dependent Variable:  Daily Time at Home (minutes)}}\\\cmidrule(r){2-6}	
			       &                    &                    &                    &                    &                    \\
High-GCD neighborhoods (Top 20\%)&      16.293$^{***}$&      19.654$^{***}$&      14.036$^{***}$&      48.501$^{***}$&      55.958$^{***}$\\
            &     (3.645)        &     (2.640)        &     (1.972)        &     (8.661)        &     (4.659)        \\
Observations&       17615        &       17615        &       17615        &       17615        &       17615        \\
R-squared   &        0.13        &        0.33        &                    &        0.07        &        0.10        \\
F-Stat      &                    &                    &                    &      433.66        &     1093.75        \\

			\vspace{0.2cm}
			&\multicolumn{5}{c}{\sc{Panel III. Dependent Variable:  Annual GHG (metric tons)}}\\\cmidrule(r){2-6}
			        &                    &                    &                    &                    &                    \\
High-GCD neighborhoods (Top 20\%)&       0.390$^{***}$&       0.223$^{***}$&       0.258$^{***}$&       1.047$^{***}$&       0.402$^{***}$\\
            &     (0.062)        &     (0.043)        &     (0.030)        &     (0.157)        &     (0.092)        \\
Observations&       17617        &       17617        &       17617        &       17617        &       17617        \\
R-squared   &        0.09        &        0.58        &                    &       -0.06        &        0.22        \\
F-Stat      &                    &                    &                    &      433.68        &     1093.77        \\
	
			\\
			\textsl{Controls:}\\
			Distance &  \checkmark & \checkmark & \checkmark & \checkmark & \checkmark\\
			Geography &  & \checkmark & \checkmark & & \checkmark \\
			State \& Metro Fixed Effects &  & \checkmark  & \checkmark & & \checkmark \\ 
			\\\bottomrule 
	\end{tabular}}
	\begin{minipage}{1\linewidth}												
		\scriptsize \textsl{Note.--- This table replicates the baseline analysis reported in Table~\ref{table: all discrete}, but aggregates all variables at the tract level instead of the census block group (CBG) level. The top 20\% of neighborhoods in the GCD distribution are classified as ``high GCD''.  Columns 1–2 report OLS estimates controlling for distance to the city center (5-km bins) and, in Column 2, additional geographic covariates including elevation, slope, ecological region, latitude, and longitude. Column 3 reports estimates from propensity score matching, using the same set of covariates as Column 2 to estimate the propensity score. Columns 4–5 replicate the same specifications using national design waves as instruments for GCD adoption. All specifications, except column 1 and 3 include metro area and state fixed effects. The coefficients with *** are significant at the 1\% confidence level; with ** are significant at the 5\% confidence level; and with * are significant at the 10\% confidence level. Robust standard errors, adjusted for clustering by county, are in parentheses. OLS and IV estimates are weighted by neighborhood population.}								
	\end{minipage}	
\end{table}	

\end{document}